\documentclass{emulateapj}

\usepackage{graphicx}
\usepackage{morefloats}
\usepackage{natbib}
\bibpunct{(}{)}{;}{a}{,}{,}

\slugcomment{To be submitted to the Astrophysical Journal}

\shorttitle{Tidal Interactions among Seyfert Galaxies: Control Experiment}
\shortauthors{Tang et al.}

\begin{document}

\title{Prevalence of Tidal Interactions Among Local Seyfert Galaxies: \\ The Control Experiment}

\author{Ya-Wen Tang}
\affil{Department of Physics, National Taiwan University, Taiwan and Institute of Astronomy \& Astrophysics, Academia Sinica, PO
Box 23-141, Taipei 10617, Taiwan} \email{ywtang@asiaa.sinica.edu.tw}

\author{Cheng-Yu Kuo}
\affil{ Institute of Astronomy \& Astrophysics, Academia Sinica, PO Box 23-141, Taipei 10617, Taiwan,
and the University of Virginia}
\email{ck2v@mail.astro.virginia.edu}

\author{Jeremy Lim}
\affil{Institute of Astronomy \& Astrophysics, Academia Sinica, PO Box 23-141, Taipei 10617, Taiwan}
\email{jlim@asiaa.sinica.edu.tw}

\and

\author{Paul T. P. Ho}
\affil{Institute of Astronomy \& Astrophysics, Academia Sinica, PO Box 23-141, Taipei 10617, Taiwan,
and the Harvard-Smithsonian Center for Astrophysics}
\email{pho@asiaa.sinica.edu.tw}

\begin{abstract}
We test whether there is a relation between the observed tidal
interactions and Seyfert activity by imaging in HI twenty inactive
galaxies at the same spatial resolution and detection threshold as
the Seyfert sample.  This control sample of inactive galaxies were
closely matched in Hubble type, range in size and inclination, and
have roughly comparable galaxy optical luminosity to the Seyfert
galaxies.  We find that only $\sim$$15\%$ of the galaxies in our
control sample are disturbed in HI, whereas the remaining
$\sim$$85\%$ show no disturbances whatsoever in HI. Even at a
spatial resolution of $\sim$10~kpc, none of the latter galaxies
show appreciable HI disturbances reminiscent of tidal features.

In a companion paper \citep{kuo08}, we report results from the
first systematic imaging survey of Seyfert galaxies in atomic
hydrogen (HI) gas.  We find that only $\sim$$28\%$ of the eighteen
Seyfert galaxies in that sample are visibly disturbed in optical
starlight.  By contrast, $\sim$$94\%$ of the same Seyfert galaxies
are disturbed spatially and usually also kinematically in HI gas
on galactic scales of $\gtrsim 20$~kpc.  In at least $\sim$$67\%$
and up to perhaps $\sim$$94\%$ of cases, the observed disturbances
can be traced to tidal interactions with neighboring galaxies
detected also in HI. The dramatic contrast between the observed
prevalence of HI disturbances in the Seyfert and control samples
implicates tidal interactions in initiating events that lead to
luminous Seyfert activity in a large fraction of local disk
galaxies.
\end{abstract}


\section{INTRODUCTION}
Active Galactic Nuclei (AGNs) are believed to be manifestations of the vigorous accretion of gas onto a central SuperMassive Black Hole (SMBH).  The origin or source of this gas, and how it is brought into the sphere of influence of the SMBH, is one of the longstanding unresolved problems in AGN research \citep[for a recent brief review, see, e.g.,][]{mart04}.  Only a very small fraction (roughly $1\%$) of galaxies in the local universe exhibit luminous AGNs, although an appreciable fraction may exhibit non-stellar nuclear activity at some level.

One of the most popular hypothesis for triggering luminous AGNs is gravitational interactions between galaxies.  Observational studies of local AGNs have mostly focussed on optically-selected samples of Seyfert galaxies.  Seyfert galaxies are preferentially found among early-tpe spiral galaxies, and exhibit very broad emission lines (from their nuclei) compared with their inactive counterparts.  Although most Seyfert galaxies do not appear to be visibly disturbed let alone interacting in optical starlight, some studies report that Seyfert galaxies more frequently possess projected or genuine neighboring galaxies (within a given angular separation) compared with matched samples of inactive galaxies \citep[e.g.,][]{sta82,dah84,raf95,dul99,kou06}.  This may constitute indirect evidence for more frequent interactions between Seyfert and their neighboring galaxies than their visual appearences would suggest.  On the other hand, there also are studies that refute reports of any differences between the frequency of Seyfert and inactive galaxies with projected neighboring galaxies \citep[e.g.,][]{mac89,fue88,der98b,sch01}.

 Studies of nearby galaxies have demonstrated that atomic hydrogen (HI) gas can reveal tidal features not visible in optical starlight \citep[e.g.,][]{yun94}.  Because the HI disk of normal spiral galaxies usually extends nearly twice as far out as the stellar optical disk, the outer regions of the HI disk is more loosely gravitationally bound.  The HI disk is therefore more susceptible to external gravitational perturbations than the stellar optical disk, and when perturbed the outskirts also take longer to dynamically relax.  This makes HI the most sensitive and enduring tracer known of gravitational interactions between galaxies.  To directly address whether Seyfert galaxies are involved in galaxy-galaxy interactions, we have therefore, for the first time, imaged uniformly a relatively large sample of Seyfert galaxies in HI gas. The full results of that study are presented in a companion paper by \citet{kuo08}.  Here, we compare the results of \citet{kuo08} with those obtained for a matched sample of inactive galaxies; i.e., a control sample.

To select a control sample, we apply the lessons learnt from optical studies that address whether galaxy-galaxy interactions are responsible for triggering Seyfert activity \citep[see, e.g., discussions in][]{fue88,der98b,dul99}.  A number of the especially earlier studies select as their control sample the closest (in projected separation) inactive galaxies with sizes comparable to the Seyfert galaxy \citep[e.g.,][]{dah84,mac89,raf95}.  As pointed by \citet{dul99}, if Seyfert galaxies lie at or close to the center of a local galaxy density enhancement, a control sample selected in this manner may then systematically lie in a region of lower galaxy density.  Consistent with this possibility, \citet{mac89} find that a larger fraction of Seyferts have projected neighboring galaxies than comparably sized inactive galaxies lying in the relatively close vicinity, but not with those selected at random across the sky.

The more recent optical studies select at random inactive galaxies with the same morphological (i.e., Hubble) type and redshift, and sometimes also size and absolute magnitude, as the Seyfert galaxies.  This matching is made either on a one-to-one basis \citep[e.g.,][]{der98a}, or over the range of properties observed \citep[e.g.,][]{dul99}.  As pointed out by both \citet{fue88} and \citet{dul99}, matching in absolute magnitudes may systematically bias the control sample to higher luminosities because of the non-negligible contribution from the AGN.  An exact match to the absolute magnitude of the Seyfert galaxy minus its AGN contribution is usually not possible as such measurements are not generally available, but is desirable to the degree possible \citep[e.g., as in][]{der98b}. By minimizing the differences in galaxy properties between the Seyfert and control samples, it is hoped that a meaningful comparison of their environments will be possible, although this is not obviously the case.  Using this method, \citet{der98b} find no significant excess of projected neighboring galaxies between their Seyfert and control samples, and at best only a marginal difference between their light asymmetries.  On the other hand, \citet[e.g.,][]{dul99} find an excess of Seyfert~2, but not Seyfert~1, with projected neighboring galaxies compared with their control sample, as was later confirmed spectroscopically by \citet{kou06}.

In our HI imaging study, we select a control sample that is matched on a one-to-one basis in Hubble type, and which range from lower to approximately the same absolute magnitude as, the sample of active galaxies selected by \citet{kuo08}.  We also matched over the observed range of optical sizes and inclinations.  No visual inspection of the control sample was made to see whether they are visibly disturbed, but we are aware that the selection criteria used may select against strongly disturbed galaxies not easily classified in Hubble type.  Indeed, using much the same criteria to select their control sample, \citet{der98a} find that a higher proportion of Seyfert galaxies are involved in late-stage mergers than their control sample, although a similar fraction of the control sample displays significant light asymmetries that could be evidence for recent interactions.  Note that only a small fraction of the active galaxies studied by \citet{kuo08} are visibly disturbed in the optical, but not so strongly that they cannot be classified in Hubble type.  Nevertheless, as we shall see, the fraction of optically-disturbed galaxies in the Seyfert sample of \citet{kuo08} is higher than that in our control sample.  Because our control sample is randomly selected from the field, their environments should be representative (or at least not biased in any peculiar way against those) of inactive galaxies with similar Hubble types, absolute magnitudes, and sizes as the Seyfert sample of \citet{kuo08}.

In \S2 we describe how we selected our control sample of inactive galaxies, and in \S3 our HI observations of these galaxies and the data reduction.  In \S4 we present the results, and in \S5 compare these results with those for the Seyfert sample studied by \citet{kuo08}.  In \S6, we provide a concise summary of our principle findings and their implications.  Unless otherwise specified, we assume throughout a Hubble constant of $H_{o} = 67 {\rm \ km \ s^{-1} \ Mpc ^{-1}}$ and $\Omega = 1$ as used by \citet{kuo08}.

\section{SAMPLE SELECTION}
As described more fully in \citet{kuo08}, the parent sample of active galaxies comprise all twenty-seven disk galaxies listed in the \citet{ver98}, plus another from the \citet[][]{ver00}, catalog at redshifts of $0.015 \leq z \leq 0.017$ and with absolute $B$-band magnitudes of $-19 \geq M_B \geq -23$ (for $H_{o} = 50 {\rm \ km \ s^{-1} \ Mpc ^{-1}}$, as used in that catalog) in the northern hemisphere ($\delta \geq 0^{\circ}$). Those galaxies were observed at an angular resolution of $\sim$60\arcsec, which at their distances of 66--75~Mpc corresponds to a spatial resolution of $\sim$20~kpc.  The latter is roughly comparable with the optical disk sizes of the active galaxies, and hence the HI observations were tailored towards the detection of structures extended on galactic scales characteristic of tidal features.

Our objective here is to conduct an equivalent HI imaging study of a comparable number of inactive galaxies matched as closely as is reasonably possible to the active galaxy sample.  We selected our control sample from the CfA Redshift Survey \citep[][see also http://cfa-www.harvard.edu/~huchra/zcat/]{huc83,huc95} according to the following criteria.

\begin{itemize}

\item[1.]  Redshift: to reduce the required observing time (which amounted to an integration time of 2~hrs and an observing time of $\sim$2.5~hrs per object in our active galaxy sample), we selected inactive galaxies at half the reshift range of our active galaxy sample; i.e., at $0.0075 \leq z \leq 0.0085$ (distance of 33.4--37.8~Mpc).  A total of 166 galaxies in the CfA redshift catalog met this criterion.

Individual samples selected in optical studies are located over a much broader range of redshifts, making one-to-one (or overall) matching in redshift (range) important for avoiding difficulties when looking for an excess of projected neighboring galaxies between two samples.  Here, our objective is to search for large-scale HI disturbances in our target galaxies, and determine whether these disturbances are produced by tidal interactions with neighboring galaxies.  In such a case, matching in redshift is not necessary so long as we observe both samples at the same spatial resolution and detection threshold (see \S3), and provided there is no significant Cosmological evolution between the two redshifts.  Of course, to look for interacting neighboring galaxies (but not necessarily to detect tidal features in the target galaxies), in both samples the field of view has to be sufficiently large to encompass any such galaxies.

\item[2.] Optical morphology and luminosity: from the parent sample of 166 inactive galaxies, we matched on a one-to-one basis the Hubble type and optical $B$-band luminosity of each active galaxy to one or more inactive galaxies.
Both the optical morphology and luminosity were taken from Leda, now renamed the Hyperleda database \citep[http://leda.univ-lyon1.fr; ][]{pat03}.  The morphology is matched to an accuracy of one Hubble subtype (e.g., for an Sa active galaxy, we matched to S0a-Sab inactive galaxies) given the inherent fuzziness in morphological classifications.  Because the absolute magnitudes of the active galaxies include a poorly determined contribution from their bright nuclei, we allowed the match in absolute magnitude to be up to one magnitude larger (i.e., optical $B$-band luminosity up to a factor of 2.5 lower) than that of the corresponding active galaxy.

\item[3.] Optical size and inclination: based once again on the Hyperleda database, we also loosely matched in optical size by restricting the galaxy diameters to the range $\sim$0\farcm8--4\farcm3 (with the majority in the range $\sim$1\farcm5--2\farcm5) as measured at the $25 {\rm \ mag \ arcsec^{-2}}$ isophotal level in the $B$-band.  This range in angular diameters is roughly comparable with the corresponding range in physical diameters of the active galaxy sample.  Because none of the active galaxies are close to face on (i.e., inclinations close to 0\degr), we also restricted the inclinations of the inactive galaxies to angles $\geq 26\degr$ with respect to the plane of the sky.

\item[4.]  Radio continuum: to avoid the effect in our dynamic range of our maps, we discarded those galaxies that lie near bright radio continuum sources detected in the NRAS VLA Sky Survey (NVSS; Condon et al. 1998) which would appear in the field of view of our HI observations. Here, we set a flux density cut at 21 cm of 1000 mJy. The detected flux densities of both sample and its companions are in the order of 100 mJy. This cut will not cause any bias when comparing results of the seyfert and control sample.

\end{itemize}

In this way, we selected twenty-seven inactive galaxies that are closely matched in Hubble type and to the degree possible absolute magnitude, {as well as range in size and inclination, to the sample of twenty-eight active galaxies in \citet{kuo08}.  The basic properties of this matched control sample are listed in Table~1.  Note that some of the values originally obtained from the Leda database have since been updated in the Hyperleda database; we list the latest, and presumably most reliable, values from the Hyperleda database.  Our selection is entirely blind to the HI content of these galaxies.

In Figure~1, we plot the number distribution in morphological types, absolute magnitudes, optical sizes, and inclinations of both the active galaxy sample in \citet{kuo08} and our control sample.  The number distribution in morphological types (top panel) shows no marked differences between the two samples within a precision of a Hubble subtype; a Kolmogorov-Smirnov (K-S) test applied to the two samples gives a K-S statistic ${\rm D}=0.18$ (converging to 0 for identical distributions) and probability ${\rm Pr}=0.74$ (1 for identical distributions).  As intended, the absolute magnitudes (second panel from top) of the control sample are skewed towards larger values (i.e., lower luminosities) than the active sample.  The distribution in optical sizes also is skewed towards somewhat smaller dimensions for the control sample, but the majority in both samples span the same range.  The control sample has more moderately inclined ($30^{\circ}$--$45^{\circ}$) than high inclined ($75^{\circ}$--$90^{\circ}$) galaxies compared to the active sample, but once again the majority in both samples span the same range.  The relatively small differences seen between the distribution of physical properties between the active and control samples should not have a significant effect on the detectability of HI tidal features in the two samples; as shown by \citet{kuo08}, the vast majority of the active galaxies, and nearly all those classified as Seyferts, exhibit HI disturbances usually in the form of tidal features irrespective of their Hubble type, absolute magnitude, optical size, or inclination within the selected range.

\section{OBSERVATIONS AND DATA REDUCTION}
We observed the control sample with the Very Large Array (VLA) of the National Radio Astronomical Observatory (NRAO).  Like the active galaxy sample studied by \citet{kuo08}, we used the most compact configuration of the VLA (the D-array).  NGC~5375 had previously been adequately (for the purpose of this experiment) observed in this configuration on 1996~August~15, and so we simply retrieved the data for this galaxy from the VLA archive.  The remaining galaxies were observed on 2003 February 19, 21, and 24, apart from NGC~5355 which was observed on 2004~July~14.  Other relevant details of the observations, including the flux, secondary (for amplitude and phase calibration), and bandpass calibrators used for each target object, are summarized in Table~2.

The correlator was configured in the same manner described in \citet{kuo08}, recording signals in orthogonal circular polarizations in sixty-four channels spanning a bandwidth of 6.25~MHz.  The corresponding channel separation in velocity is $\sim$$21.2 {\rm \ km \ s^{-1}}$ (the actual velocity resolution is $\sim$$25.2 {\rm \ km \ s^{-1}}$), and altogether span a velocity range of $\sim$$1350 {\rm \ km \ s^{-1}}$.  For the assumed Cosmology, this velocity range corresponds to a redshift interval of ${\Delta}z \approx 0.0045$ (distance range of $\sim$20~Mpc), compared with the redshift interval of ${\Delta}z \approx 0.001$ for the control sample.  The central channel is set to the systemic heliocentric velocity of the HI line if previously measured (usually with a single-dish telescope), or if not then at its reported optical redshift.

At 21~cm, the primary beam of the VLA is $\sim$32\arcmin\ at full-width half-maximum, corresponding to a diameter of $\sim$330~kpc at the distance of our control sample.  Our field of view for the control sample is, of course, only half the linear diameter of our field of view for the active galaxy sample, which is located at twice the distance.  As described in \citet{kuo08}, the projected separation between the Seyfert and their interacting neighboring galaxies is  mostly (for $\sim$$85\%$ of the sample) $\lesssim 100$~kpc. Our field of view is therefore sufficiently large to encompass most of the interacting neighboring galaxies detected around the Seyfert galaxies studied by \citet{kuo08} even if the latter was placed at the distance of our control sample.

We targeted the same detection threshold at the same spatial resolution as \citet{kuo08} for their active galaxy sample.  Because the control sample is a factor of two closer in distance, their HI intensity is a factor of 4 higher for the same gas mass.  Everything else being equal, this would require a factor of 16 shorter integration time for our control sample, a considerable saving given that the integration time for each active galaxy in \citet{kuo08} was $\sim$2~hrs.  In practise, to achieve the same spatial resolution of $\sim$20~kpc (i.e., angular resolution of $\sim$120\arcsec) as in the active galaxy sample (which was imaged at an angular resolution of $\sim$60\arcsec), we had to discard baselines that were more than one-third the length of the longest baselines in the active galaxy sample (i.e., using baseline up to only $1.5 \rm \ k\lambda$).  Because there are more shorter than longer baselines, the actual integration time per object in our control sample was therefore $\sim$12~mins (still a considerable saving in time over 2~hrs) and the observing time per object $\sim$20~mins (compared with $\sim$2.5~hrs for the active galaxy sample).  In this way, we achieved a rms flux density of $\sim$$1.8 {\rm \ mJy\ beam^{-1}}$ in one channel, corresponding to a 5$\sigma$ detection threshold in HI gas column density of $\sim$$3 \times 10^{19} {\rm \ cm^{-2}}$ per synthesized beam of diameter $\sim$120\arcsec\ (20~kpc) in a given channel with a velocity width $\sim$$25.2 {\rm \ km \ s^{-1}}$.  This is the same detection threshold as reached by \citet{kuo08} in their active galaxy sample.

Like \citet{kuo08}, we performed all the data reduction in the standard manner using the NRAO AIPS package.  Unlike for the active galaxy sample, we found no radio frequency interference (RFI) in the data for our control sample, presumably because their redshifted HI lines lie closer to the protected band at 21~cm.  Following bandpass calibration, as well as time-dependent amplitude and phase calibration, we subtracted the continuum emission from the visibility data by interpolating between line-free channels (determined through trial and error) on either side of the HI line.  For NGC~4256, where there is a relatively strong continuum source in the field, we first subtracted the continuum emission of that source in the visibility plane.  Finally, we made maps of each channel, and corrected for the primary beam response of the antennas.  We also combined the channel maps (before primary beam correction) to make maps in total intensity (zeroth moment) and intensity-weighted mean velocity (first moment), where we tried several different combinations of spectral smoothing and minimum signal cutoff level in each channel so as to suppress the noise and bring out faint diffuse features.

\section{Results}
We detected twenty-one of the twenty-seven galaxies in our control sample.  Two of the six galaxies not detected, NGC~4128 and NGC~4543, have (to the best of our knowledge) not previously been observed in HI.  The reported upper limits in the integrated HI intensities for NGC~4578 (Hutchmeier et al. 1998) and NGC~5326 (Theureau et al. 1998) are below that attained in our observations.  NGC~3468 has a reported integrated HI intensity of $1.40 {\rm \ Jy \ km \ s^{-1}}$ (with no reported uncertainty) in \citet{Huchtmeier1989}, and should therefore have been detected in our observation.  On checking the original source of the detection in \citet{Giovanelli1981}, we found that the actual galaxy detected was UGC~3468 and not NGC~3468 as reported in \citet{Huchtmeier1989}.  NGC~5355 has a reported integrated HI intensity of $14.6 \pm 1.8 {\rm \ Jy \ km \ s^{-1}}$ in \citet{Huchtmeier1989}, and should therefore have been detected in our observation.  On checking the original source of the detection, Richter and Huchtmeier (1991) mentioned that the 8\farcm8 beam of the Effelsburg telescope used for this observation would include also NGC~5353 and NGC~5354.  We note that this beam also includes NGC~5350, which is one of the galaxies in our control sample.  We detected NGC~5350 at a much higher integrated HI intensity than that reported by Richter and Huchtmeier (1991) for NGC~5355, suggesting that the actual HI detection was of NGC~5350.

The twenty-one galaxies that we detected in HI comprise our ensemble control sample from which we draw statistical results.  For comparison, there are twenty-three active galaxies in the ensemble sample of \citet{kuo08}, of which eighteen are classified as Seyfert galaxies. The HI maps of the ensemble control sample are shown in Figures~2--25, and their HI properties summarized in Table~3.  We have overlaid each integrated HI intensity (zeroth moment) map on an optical (either the red or blue filters) image from the 2nd Digitized Sky Survey (DSS2).  This database provides a relatively uniform set of optical images to search for any disturbances in optical starlight.

We also detected in HI a number of galaxies in the same field of view as our target galaxies.  Those galaxies lying within the primary beam (i.e., within a radius of $\sim$15\farcm8 about phase center) are listed in Table~4, and are referred to by their names from optical catalogs.  We also list their optical redshifts where available, as well as their corrected apparent $B$-band magnitudes and Hubble types as listed in the Hyperleda database.  The measured HI properties of these neighboring galaxies are summarized in Table~5.  Apart from neighboring galaxies detected in HI, we also searched the Hyperleda database in a radius of $\sim$15\arcmin\ about each galaxy in our control sample to obtain a measure of the richness of their fields.  We emphasize that this is by no means an exhaustive search.

\subsection{Individual Galaxies}
We present here the results for each galaxy, highlighting features of most relevance to this work.  As in \citet{kuo08}, we separate the individual galaxies in our ensemble control sample into the following three groups.

\begin{itemize}
\item[1.]  Those that clearly show HI tidal features tracing interactions with neighboring galaxies.  These features are in the form of tidal bridges that connect the two interacting galaxies, or an extension from one or both galaxies in the direction of the other (i.e., an incomplete tidal bridge).  In addition, features in the form of tidal tails comprising a protrusion or curved extension from one galaxy on the side away from the other interacting galaxy may be seen.  Only one of the twenty-one galaxies in our ensemble control sample of inactive galaxies fall into this group.  By comparison, thirteen of the twenty-three galaxies in the ensemble sample of active galaxies, including twelve of the eighteen classified as Seyferts, fall into this group.

\item[2.] Those that clearly show both spatial and kinematic disturbances in HI, but which cannot be directly linked to interactions with neighboring galaxies if any.  These disturbances are in general less prominent than the tidal features observed in the first group.  Only two of the twenty-one galaxies in our ensemble control sample of inactive galaxies fall into this group.  Four of the twenty-three galaxies in the ensemble sample of active galaxies, including three of the eighteen classified as Seyferts, fall into this group.

\item[3.] Those that show marginal or no detectable HI disturbances.  Eighteen of the twenty-one galaxies in our ensemble control sample of inactive galaxies fall into this group.  Only one exhibits weak HI disturbances; the remaining seventeen exhibit no discernible HI disturbances.  By comparison, only six of the twenty-three galaxies in the ensemble sample of active galaxies, including only three (two of which exhibit weak HI disturbances) of the eighteen classified as Seyferts, fall into this group.

\end{itemize}

Because the vast majority of the galaxies in our control sample exhibit no detectable HI disturbances at an angular resolution of $\sim$120\arcsec\ ($\sim$20~kpc), for these undisturbed galaxies we show only HI maps at the full angular resolution attained in our observations of $\sim$60\arcsec\ ($\sim$10~kpc).  The maps at $\sim$120\arcsec\ angular resolution contain no extra information.  For the disturbed galaxies, moment maps are shown at angular resolutions of both $\sim$120\arcsec\ and $\sim$60\arcsec\ for the reader to compare, and HI channel maps at an angular resolution of $\sim$120\arcsec\ only. The HI channel maps of the undisturbed galaxies are available in the electronic version of this paper (Figs.~28--44).(For astroph readers, the channel maps are available at the link http://www.astro.virginia.edu/~ck2v/maps) At an angular resolution of $\sim$60\arcsec, the corresponding 5$\sigma$ detection threshold in HI gas column density is $\sim$$7 \times 10^{19} {\rm \ cm^{-2}}$ per synthesized beam in a given channel with a velocity width $\sim$$25.2 {\rm \ km \ s^{-1}}$, which is higher than the Seyfert sample. Because the Seyfert sample is twice further away from us comparing with the control sample, the detection limit in terms of the HI mass of the control sample ($\sim$ 6 $\times 10^{7} M_{\sun}$) is lower than of the seyfert sample ($\sim$ 7 $\times 10^{8} M_{\sun}$).

\subsubsection{Group I: HI tidal features tracing interactions with neighboring galaxies}
At a spatial resolution of $\sim$20~kpc only one of our ensemble sample of inactive galaxies exhibit tentative HI disturbances, which only in the map at a spatial resolution of 10~kpc can be clearly seen to trace to tidal interactions with neighboring galaxies.  Nevertheless, we generously place this sole example of its kind in Group~1.  When categorizing the ensemble sample of active galaxies in \citet{kuo08}, we took a more conservative approach.\newline

\hspace{3 cm}
NGC~2543
\newline

NGC~2543 is a barred Sb galaxy (all Hubble types from the Hyperleda database, just as for the active galaxies in our companion paper). It appears somewhat asymmetric in the optical, although not obviously disturbed. A relatively dim galaxy at a comparable optical redshift, PGC~80408, lies $\sim$3\farcm4 ($\sim$36~kpc) south of NGC~2543.

Our HI image at a spatial resolution of $\sim$20~kpc (Fig.~2)
reveals that NGC~2543 shares a common envelope with PGC~80408.  An
examination of the channel maps at this spatial resolution
(Fig.~3) reveals tentative evidence for a tidal bridge (which, if
real, is at best marginally spatially resolved) connecting the two
galaxies.  Our HI image at a spatial resolution of $\sim$10~kpc
(also shown in Fig.~2), and also the channel maps at this angular
resolution (also Fig.~3), clearly reveals that the connection
between the two galaxies is indeed a tidal bridge.  In addition,
the HI disk of NGC~2543 is clearly disturbed on its south-western
side, which on this side extends much further than on the
north-eastern side and curls in a direction towards its
interacting neighbour PGC~80408.

\subsubsection{Group II: HI disturbances likely produced by interactions}
Two of the galaxies in our ensemble control sample exhibit spatial and kinematic disturbances in HI, but unlike the situation in Group~I the observed perturbations cannot be immediately linked to interactions with neighboring galaxies.\newline

\hspace{3cm}
NGC~2551
\newline

NGC~2551 is a Sa galaxy.  A dimmer
galaxy at a comparable optical redshift, UGC~4390, lies
$\sim$14\farcm4 ($\sim$146~kpc) to the north-east of NGC~2551.
Neither galaxies appear to be visibly disturbed in the optical.

We detected both NGC~2551 and UGC~4390 in HI, as well as the
relatively faint galaxy PGC~2755603, which has no previously
reported optical redshift (Figs.~4 and 5). At a spatial resolution
of $\sim$20~kpc (Fig.~5), the HI disk of NGC~2551 appears to be
disturbed, with this disturbance especially prominent in the first
moment map (kinematics).  The observed disturbance in the HI disk
is even more prominent at an angular resolution of $\sim$10~kpc
(Fig.~4), but not easily apparent in the channel maps (Fig.~6). By
comparison, UGC~4390 does not appear to be disturbed in HI at
either spatial resolutions, although it is possible that the HI
kinematic axis is not aligned with the optical major axis (the
optical major axis is poorly defined in the outer regions of this
galaxy).  PGC~2755603 was not spatially resolved in HI.  It is not
clear whether the observed disturbance in NGC~2551 was caused by
tidal interactions with (one of) these or other galaxies.\newline

\hspace{3cm}
NGC~3094
\newline

 NGC~3094 is a barred Sa galaxy that
does not appear to be obviously disturbed in the optical. In HI
(Fig.~7), we detected NGC~3094 and also PGC~2806972, a relatively
faint galaxy that liest $\sim$7\farcm5 ($\sim$78~kpc) to the east
of NGC~3094. PGC~2806972 has no previously reported optical
redshift.

NGC~3094 exhibits a HI protrusion on the south-eastern side that
is kinematically discontinuous from its HI disk.  A close
examination of the channel maps (Fig.~8) reveals that this
protrusion in physically linked to NGC~3904. These channel maps
also reveal prominent north-south extensions over a narrow
velocity range not seen in the moment maps; these extensions are
orthogonal to the HI kinematic axis of the galaxy.  There are no
cataloged galaxies at the location of this protrusion, nor any
optical counterpart visible in the DSS2 image.  It is possible
that this protrusion is an incomplete tidal bridge caused by
interactions with PGC~2806972.

\subsubsection{Group III: Weak or no HI disturbances}
Eighteen of the twenty-one galaxies fall into this group,
accounting for the vast majority of the galaxies in our ensemble
control sample.  Of these eighteen, seventeen do not show any
evidence whatsover for disturbances in HI.  Only NGC~3976, which
as we explain below has recently been classified as an active
galaxy, shows evidence for weak disturbances.\newline

\hspace{3cm}
NGC~3976
\newline

NGC~3976 is a barred Sb galaxy. 
It is clearly disturbed in the optical (Fig.~9), exhibiting
asymmetric extensions on the north-eastern and south-western sides
of the disk. Although NGC~3976 is not classified as an active
galaxy by \citet{ver98}, the source used to select the active
galaxy sample, it is classified as a Seyfert~2 with $M_B=-21.1$ in
the latest version of this catalog \citep{ver06}.  Thus, if
NGC~3976 was placed at the same range of redshifts as the active
galaxy sample, it would have been selected as part of that sample.
For internal consistency, however, we retain NGC~3976 in our
control sample.  If included in the active galaxy sample, this
would make the conclusions reached in our study only stronger.  A
relatively faint galaxy at a comparable optical redshift,
PGC~37490, lies $\sim$4\farcm5 to the south of NGC~3976.

We detected NGC~3976, but not PGC~37490.  At a spatial resolution
of $\sim$20~kpc, the HI disk of NGC~3976 appears to be asymmetric
(Fig.~9).  The HI gas extends further out on the north-eastern
side compared to the south-western side, giving the disk a
lopsided appearence in HI.  In addition, the HI kinematic axis
appears to lie at a different position angle compared with the
major axis of the optical disk.  At a spatial resolution of
$\sim$10~kpc (also shown in Fig.~9), this asymmetric appearence
appears to be caused in part (but perhaps not wholly) by the
larger HI extension of the north-eastern compared with the
south-western spiral arm.\newline

\hspace{3cm}
IC~2461
\newline

 IC~2461 is an edge-on Sb galaxy that
does not appear to be visibly disturbed in the optical.  There are
no cataloged galaxies at a comparable optical redshift within
15\arcmin\ of the galaxy.

At a spatial resolution of $\sim$20~kpc, the HI disk of IC~2461
appears to be symmetric both spatially and kinematically with no
evidence for perturbations.  At a spatial resolution of
$\sim$10~kpc (Fig.~10), the HI major and kinematic axes appears to
exhibit an aticlockwise twist on the north-western side of the
disk, perhaps indicative of a warp.\newline

\hspace{3cm}
NGC~3835
\newline

NGC~3835 is a Sab galaxy that is not
visibly disturbed in the optical.  There are no cataloged galaxies
at a comparable optical redshift within 15\arcmin\ of the galaxy.

At a spatial resolution of $\sim$20~kpc, the HI disk of NGC~3835
appears to be symmetric both spatially and kinematically with no
evidence for perturbations.   The same also is true at a spatial
resolution of $\sim$10~kpc (Fig.~11), where the central inward
contraction of the intensity contours in the zeroth moment maps is
probably caused by a depression in HI intensity at the center of
the galaxy (as is often observed in nearly disk galaxies).\newline

\hspace{3cm}
NGC~4067
\newline

 NGC~4067 is a Sb galaxy that is not
visibly disturbed in the optical.  There are no cataloged galaxies
at a comparable optical redshift within 15\arcmin\ of the galaxy.

At a spatial resolution of $\sim$20~kpc, the HI disk of NGC~4067
appears to be symmetric both spatially and kinematically with no
evidence for perturbations.   At a spatial resolution of
$\sim$10~kpc (Fig.~12), the HI disk appears to be more prominent
or extended on the south-western side compared with the
north-eastern side.  Other than this small asymmetry, the HI disk
does not exhibit any spatial nor kinematic disturbances.  Once
again, the central inward contraction of the intensity contours in
the zeroth moment maps is probably caused by a depression in HI
intensity at the center of the galaxy.\newline

\hspace{3cm}
NGC~4256
\newline

 NGC~4256 is an edge-on Sb galaxy that
is not visibly disturbed in the optical.  There are five
relatively faint cataloged galaxies at a comparable optical
redshift within 15\arcmin\ of NGC~4256.  Their separations range
from 0\farcm4 ($\sim$4~kpc) to 14\arcmin\ ($\sim$153~kpc) from
NGC~4256.  None of these neighboring galaxies were detected in HI.

The HI gas mass of NGC~4256 is the lowest among all of the
galaxies in our ensemble control sample, and hence the
corresponding signal-to-noise ratio of the HI map for this galaxy
poorest.  At a spatial resolution of $\sim$20~kpc, the HI disk of
NGC~4256 appears to be symmetric both spatially and kinematically
with no evidence for perturbations.  The same also is true at a
spatial resolution of $\sim$10~kpc (Fig.~13).\newline

\hspace{3cm}
NGC~4275
\newline

 NGC~4275 is a Sb galaxy that has the
smallest inclination among all of the galaxies in our ensemble
control sample.  There are no cataloged galaxies at a comparable
optical redshift within 15\arcmin\ of the galaxy.

At a spatial resolution of $\sim$20~kpc, the HI disk of NGC~4275
appears to be symmetric both spatially and kinematically with no
evidence for perturbations. The same also is true at a
spatial resolution of $\sim$10~kpc (Fig.~14).\newline

\hspace{3 cm}
NGC~4351
\newline

 NGC~4351 is a barred Sab type galaxy
that appears to be disturbed in the sense that it is lopsided
in the optical, being more extended on the western than eastern
side of the galaxy. There are no cataloged galaxies at a
comparable optical redshift within 15\arcmin\ of the galaxy.

At a spatial resolution of $\sim$20~kpc, the HI disk of NGC~4351
appears to be symmetric both spatially and kinematically with no
evidence for perturbations.   At a spatial resolution of
$\sim$10~kpc (Fig.~15), the HI disk is asymmetric in that it is
more prominent or extended on the western side, just like in
the optical.\newline

\hspace{3cm}
NGC~4384
\newline

NGC~4384 is a Sa galaxy that is not
visibly disturbed in the optical.  There is a relatively faint
galaxy, PGC~2478952, with a comparable optical redshift at
$\sim$14\farcm9 ($\sim$162~kpc) from NGC~4384.  That galaxy was
not detected in HI.

At a spatial resolution of $\sim$20~kpc, the HI disk of NGC~4384
appears to be symmetric both spatially and kinematically with no
evidence for perturbations.   At a spatial resolution of
$\sim$10~kpc (Fig.~16), the HI disk may be somewhat asymmetric in
that the eastern side appears to be more prominent or extended.\newline

\hspace{3 cm}
NGC~4470
\newline

 NGC~4470 is a Sa type galaxy that is
not visibly disturbed in the optical.  The galaxy that is closest
in optical redshift and which lies within 15\arcmin\ of NGC~4470
differs in systemic velocity by $\sim$$900 {\rm \ km \ s^{-1}}$,
and lies outside the velocity coverage of our HI observation.

At a spatial resolution of $\sim$20~kpc, the HI disk of NGC~4470
appears to be symmetric both spatially and kinematically with no
evidence for perturbations.   At a spatial resolution of
$\sim$10~kpc, the HI disk may show weak asymmetries
(Fig.~17).\newline

\hspace{3cm}
NGC~4513
\newline

NGC~4513 is a S0 galaxy that is not
visibly disturbed in the optical.  There are no cataloged galaxies
at a comparable optical redshift within 15\arcmin\ of the galaxy.

At a spatial resolution of $\sim$20~kpc, the HI disk of NGC~4513
appears to be symmetric both spatially and kinematically with no
evidence for perturbations.   The same also is true at a spatial
resolution of $\sim$10~kpc (Fig.~18).  This galaxy has the second
lowest HI gas mass among our ensemble control sample, with the
resulting low signal-to-noise ratio making the presumed central
depression in HI gas particularly prominent.\newline

\hspace{1.7cm}
NGC~4567 and NGC 4568
\newline

 Both NGC~4567 and
NGC~4568 are Sbc galaxies that form an optically-overlapping pair
with comparable optical redshifts.  In this way, NGC~4567/NGC~4568
resemble UGC~3995/UGC~3995A in the active galaxy sample.  Both
NGC~4567 and NGC~4568 are included in our control sample.  There
are no other cataloged galaxies at a comparable optical redshift
within 15\arcmin\ of the galaxy.

Despite their apparent proximity and similar redshifts, as in the
case of UGC~3995/UGC~3995A we detect no HI disturbances at a
spatial resolution of $\sim$20~kpc, nor at a spatial resolution of
$\sim$10~kpc (Fig.~19), in either NGC~4567 or NGC~4568.  Indeed,
in a much deeper HI observation at an angular resolution of
$\sim$20\arcsec, \citet{Iono2005} detected no extended tidal
features in this pair of galaxies.\newline

\hspace{3 cm}
NGC~4591
\newline

 NGC~4591 is a Sb galaxy that is not
visibly disturbed in the optical.  There are no cataloged galaxies
at a comparable optical redshift within 15\arcmin\ of the galaxy.

At a spatial resolution of $\sim$20~kpc, the HI disk of NGC~4591
appears to be symmetric both spatially and kinematically with no
evidence for perturbations.   The same also is true at a spatial
resolution of $\sim$10~kpc (Fig.~20).\newline

\hspace{3cm}
NGC~4814
\newline

 NGC~4814 is a Sb galaxy that is not
visibly disturbed in the optical.  There is a relatively faint
galaxy, SDSSJ~125426.85+582348.9, with a comparable optical
redshift at $\sim$7\farcm9 ($\sim$86~kpc) from NGC~4384.  That
galaxy was not detected in HI.

At a spatial resolution of $\sim$20~kpc, the HI disk of NGC~4814
appears to be symmetric both spatially and kinematically with no
evidence for perturbations.  The same also is true at a spatial
resolution of $\sim$10~kpc (Fig.~21).\newline

\hspace{3cm}
NGC~4964
\newline
 NGC~4964 is a S0-a galaxy that is not
visibly disturbed in the optical. There are no cataloged galaxies
at a comparable optical redshift within 15\arcmin\ of the galaxy.

At a spatial resolution of $\sim$20~kpc, the HI disk of NGC~4964
appears to be symmetric both spatially and kinematically with no
evidence for perturbations.  The same also is true at a spatial
resolution of $\sim$10~kpc (Fig.~22).\newline

\hspace{3cm}
NGC~5289
\newline

NGC~5289 is a nearly edge-on barred Sa
galaxy that is not visibly disturbed in the optical.  There are
six mostly relatively faint cataloged galaxies at a comparable
optical redshift within 15\arcmin\ of NGC~5289.  Their separations
range from 5\farcm3 ($\sim$58~kpc) to 12\farcm7 ($\sim$138~kpc)
from NGC~5289.  Only of these neighboring galaxies, NGC~5290,
lying furthest away from and brighter in the optical than
NGC~5289, was detected in HI.

At a spatial resolution of $\sim$20~kpc, the HI disk of NGC~5289
appears to be symmetric both spatially and kinematically with no
evidence for perturbations.  The same also is true at a spatial
resolution of $\sim$10~kpc (Fig.~23).\newline

\hspace{3cm}
NGC~5350
\newline

 NGC~5350 is a barred Sbc galaxy that
appears to be somewhat asymmetric although not obviously disturbed
in the optical.  There are seven cataloged galaxies at a
comparable optical redshift within 15\arcmin\ of NGC~5350, two of
which are comparably bright and the remainder relatively faint.
None of these neighboring galaxies were detected in HI.

At a spatial resolution of $\sim$20~kpc, the HI disk of NGC~5350
appears to be symmetric both spatially and kinematically with no
evidence for perturbations.  The same also is true at a spatial
resolution of $\sim$10~kpc (Fig.~24).\newline

\hspace{3cm}
NGC~5375
\newline

 NGC~5375 is a barred Sab galaxy that
is not visibly disturbed in the optical.  There is a relatively
faint galaxy, PGC~49623, with a comparable optical redshift at
$\sim$4\farcm6 ($\sim$47~kpc) from NGC~5375.  That galaxy was not
detected in HI.  Instead, we detected the galaxy MAPS-NGP
O-325-1536477 (as cataloged in the NASA Extragalactic Databse),
which has no previously reported optical redshift, which is
9$\farcm$4 ($\sim$131 kpc) to the south-east of NGC~5375.

At a spatial resolution of $\sim$20~kpc, the HI disk of NGC~5375
appears to be symmetric both spatially and kinematically with no
evidence for perturbations.  The same also is true at a spatial
resolution of $\sim$10~kpc (Fig.~25).

\subsection{Other Galaxies in Target Fields}
Apart from the optically-overlapping pair NGC~4567/NGC~4568, we
detected only six other galaxies in HI within the primary
beam of our target fields. None of these galaxies exhibit HI
disturbances at either 20~kpc or 10~kpc spatial resolution.  The
HI maps of these galaxies are shown together with the maps for the
control sample galaxies.

\subsection{Ensemble Statistics}
We now make statistical inferences from the results based on our
ensemble control sample of twenty-one galaxies that we mapped in
HI.  Only four of the twenty-one ($\sim$$19\%$) galaxies in the
control sample exhibit detectable disturbances at a spatial
resolution of $\sim$20~kpc. By contrast, seventeen of the
twenty-one ($\sim$$81\%$) galaxies exhibit no HI disturbances
whatsoever at this spatial resolution. One of these disturbed
galaxies (NGC~3976) has since been found to be a Seyfert~2 galaxy,
and so only three of the twenty ($15\%$) inactive galaxies are
actually disturbed in HI, whereas seventeen of twenty
($\sim$$85\%$) show no HI disturbances.  The HI disturbances be
directly traced to tidal interactions with neighboring galaxies in
only one case.  In the remaining two cases with HI disturbances,
the observed disturbances are likely to be produced by tidal
interactions with neighboring galaxies also detected in HI.  By
comparison, seventeen of the eighteen ($\sim$$94\%$) Seyfert
galaxies in the ensemble sample of \citet{kuo08} exhibit HI
disturbances, twelve ($\sim$$67\%$) of which are interacting with
neighboring galaxies (Group~I), another three ($\sim$$17\%$)
probably interacting with neighboring galaxies (Group~II), and the
remaining two ($\sim$$12\%$) weakly disturbed (Group~III).

Only two of the twenty-one ($\sim 10\%$) galaxies in
the control sample are visibly disturbed in optical DSS2 images.
Leaving aside NGC~3976 since found to be a Seyfert~2 galaxy, only
one of twenty ($\sim 5\%$) inactive galaxies are optically
disturbed.  By comparison, six of the twenty-three ($\sim$$26\%$)
galaxies in the ensemble sample of active galaxies studied by
\citet{kuo08} are visibly disturbed in optical DSS2 images,
including five of the eighteen galaxies ($\sim$$28\%$) classified
as Seyferts.  As pointed out in \S~1, the selection criteria used
may select against (strongly) optically disturbed galaxies in the
control sample.  Nonetheless, the relatively small difference in
the frequency of optical disturbances between the Seyfert and
control samples ($\sim$$28\%$ vs $\sim$$5\%$) pales besides the
dramatic difference in the frequency of HI disturbances between
these two samples ($\sim$94$\%$ vs $\sim$$15\%$).

In case of any severe selection bias against optically-disturbed galaxies in our control sample, here we compile statistics for just the optically-undisturbed galaxies in both the Seyfert and control sample.  Of the thirteen such Seyfert galaxies, twelve ($\sim$$92\%$) exhibit HI disturbances, nine ($\sim$$69\%$) of which are interacting with neighboring galaxies (Group~I), one ($\sim$$7\%$) possibly interacting with a neighboring galaxy (Group~II), and two ($\sim$$15\%$) weakly disturbed (Group~III).  By comparison, only three of the twenty ($\sim$$15\%$) optically-undisturbed inactive galaxies are disturbed in HI.

\section{INTERPRETATION AND DISCUSSION}

\subsection{Relative HI Gas Masses}
We first examine whether the dramatic difference in HI disturbances between the active and control samples might be caused by differences in their HI gas masses.  This could happen if, for example, the same fractional HI gas mass is displaced into tidal features of equal spatial dimensions in a given interaction.  In such a case, interacting galaxies with lower HI gas masses will exhibit dimmer tidal features, which may fall below the detection threshold in a given observation.

In Figure~26, we plot the HI gas masses of the active and control samples.  As can be seen, the active sample is distributed towards somewhat higher HI gas masses than the control sample.
A K-S test gives a statistic ${\rm D}=0.30$ (converging to 0 for identical distributions) and probability ${\rm Pr}=0.23$ (1 for identical distributions).  More importantly, most of the galaxies in our control sample span the same range of HI gas masses as the active sample for which we detect HI disturbances. We therefore believe that the dramatic difference in HI disturbances between the two samples cannot be caused by the relatively small difference in their overall distributions of HI gas masses.  The possibility that tidal features in our control sample somehow conspire to be less spatially extended than those in the active sample is ruled out by our HI images of the control sample at a spatial resolution of $\sim$10~kpc.

\subsection{Incidence of Neighboring Galaxies}
In Figure~27, we show the cumulative fraction of Seyfert galaxies (solid line) studied by \citet{kuo08} with (candidate) interacting neighboring galaxies (i.e., in Groups~I, II, or III), plotted as a function of their projected separations (for the few with multiple interacting neighboring galaxies, their nearest such neighbor).  As explained in \citet{kuo08}, over the range of projected separations plotted (up to 90~kpc), there are only two neighboring galaxies not identified to be (possibly) interacting with their respective (the same) Seyfert galaxy, the third of which is identified as the interacting neighbor.  Thus, including all possible neighboring galaxies around Seyfert galaxies do not change this figure, at least within the range of projected separations plotted.  In the same figure, we also plot the cumulative fraction of galaxies in our control sample (dashed line) with neighboring galaxies within the same range of projected separations.  (Keep in mind that our observations span the same velocity interval for both samples.)  Our detection threshold is relatively uniform (within $\sim$$20\%$) and essentially the same for both the Seyfert and control samples at projected separations up to $\sim$90~kpc.  As can be seen, the frequency of Seyfert with neighboring galaxies is clearly much higher than that of the control sample over the range of projected separations plotted.  This difference simply reflects the prevalence of tidal interactions in the Seyfert but not control sample.

As mentioned in \S1, some optical studies find an excess of Seyfert with (projected) neighboring galaxies compared with inactive galaxies, whereas others do not.  A discussion of the different methodologies used in different studies, as well as reasons for their conflicting results, is beyond the scope of this paper.  Any comparison made between previous optical and our HI studies should keep in mind that our HI imaging study is not designed to address differences between the fraction of Seyfert and inactive galaxies with neighboring galaxies.  (Rather than looking for any such difference, we directly determine what fraction of Seyfert and inactive galaxies show HI disturbances, and where possible determine whether these disturbances are caused by tidal interactions with neighboring galaxies.)  Both the study of \citet{kuo08} and that reported here can only detect relatively gas-rich galaxies, and so a HI census of neighboring galaxies is likely to be less complete than an optical census.

Nevertheless, to see what can be learnt, we compare the results of our HI studies with the optical studies of \citet{dul99} and \citet{kou06}, who measured the fraction of galaxies in both their Seyfert and control samples with (projected) neighboring galaxies as a function of their projected separations.  Both these studies employed the same Seyfert and control samples; the Seyfert sample was taken from the catalog of \citet{lip88}, which contained most if not all of the then known Seyfert galaxies, and the matched control sample from the CfA survey \citep{huc83}.

\citet{dul99} searched for projected neighboring galaxies within a radius of $\sim$140~kpc using the (first) Digitized Sky Survey (DSS).  They found that in both their Seyfert and control samples, the fraction having projected neighboring galaxies with diameters 4--10~kpc increases as the search radius widens, reaching between $80\%$--$100\%$ within a search radius of $\sim$100~kpc \citep[see Fig.~1 of][]{dul99}.  At any given search radius, the fraction of Seyfert~1 and inactive galaxies possessing projected neighboring galaxies is comparable.  On the other hand, there is a relatively large difference between the fraction of Seyfert~2 and inactive galaxies possessing projected neighboring galaxies at the smallest search radius of 10~kpc, with this difference diminishing as the search radius widens until disappearing at $\sim$100~kpc.  Thus, the difference, if any, between the fraction of Seyfert~2 and matched samples of inactive field galaxies possessing projected neighboring galaxies depends at least in part on the search radius used.  We note that this may explain the seemingly contradictry results between some optical studies; i.e., those using relatively small search radii finding a higher fraction of Seyfert with projected neighboring galaxies than in matched samples of inactive galaxies, whereas those using relatively large search radii finding no appreciable difference.

Our results agree with those of \citet{dul99} in two important respects.  First, nearly all the Seyfert galaxies studied by \citet{dul99} have projected neighboring galaxies within $\sim$100~kpc.  Second, we both see an excess of Seyfert (our ensemble sample contains a much larger fraction of Seyfert~2 than Seyfert~1) having (projected) neighboring galaxies at projected separations smaller than $\sim$90~kpc compared with the control sample.  Where we differ is in our seeing this excess becoming increasingly larger with projected separation up to 90~kpc, whereas \citet{dul99} find this excess to become increasingly smaller with projected separation before disappearing at $\sim$100~kpc.  Keep in mind, however, that our HI imaging observations cleanly pick out genuine neighboring but only gas-rich galaxies, whereas those of \citet{dul99} may include spurious objects.

\citet{kou06} repeated the study of \citet{dul99} but now armed with redshifts from the second CfA as well as Southern Sky Redshift Surveys.  Confirming the trends seen by \citet{dul99}, \citet{kou06} find a relatively large difference between the fraction of Seyfert~2 (but, again, not Seyfert~1) with neighboring galaxies within a velocity difference of $200 {\rm \ km \ s^{-1}}$ compared with inactive galaxies at the smallest measured projected separation of 7~kpc, with this difference diminishing as the search radius widens until disappearing at roughly 70~kpc (scaled to $H_{o} = 67 {\rm \ km \ s^{-1} \ Mpc ^{-1}}$).  With a maximum search radius of $\sim$70~kpc and a limiting magnitude of $m_B \sim 15.5$ (1 magnitude brighter than the LMC at the distance of our sample), this study would have picked up only a small fraction of the interacting neighboring galaxies that \citet{kuo08} detected around their Seyfert sample.  Indeed, \citet{kou06} find that only $\sim$$10\%$ of Seyfert~1 and $\sim$$30\%$ of Seyfert~2 galaxies have neighboring galaxies within the range of parameters searched.  Going to a lower limiting magnitude of $m_B \sim 18.5$ (corresponding to the SMC at the distance of our sample) for a subset of their sample, \citet{kou06} find that about twice as many still of both Seyfert~1 and Seyfert~2 galaxies have neighboring galaxies within a projected separation of $\sim$50~kpc.  This fraction is comparable to that we found in our Seyfert sample ($\sim$$40\%$ at a projected separation of $\sim$50~kpc).  Unfortunately, \citet{kou06} did not search for neighboring galaxies down to the the same low limiting magnitudes around their control samples, and hence a more detailed comparison cannot be made.

\subsection{Implicating Tidal Interactions in Triggering AGNs}
The contrast between the prevalence of HI disturbances in the active and control samples is dramatic.  In the active galaxy sample, we can only find a few galaxies that are not disturbed spatially and usually also kinematically on galactic-wide ($\gtrsim 20$~kpc) scales.  In the control sample, we find the opposite result, with most exhibiting no detectable spatial or kinematic disturbances on both the same as well as smaller ($\gtrsim 10$~kpc) scales.

Specifically, of the eighteen galaxies in the ensemble sample of twenty-three active galaxies classified as Seyferts, seventeen ($\sim$$94\%$) exhibit HI disturbances.  In at least $\sim$$67\%$ and possibly as high as $\sim$$94\%$ of cases, the observed HI disturbances can be traced to tidal interactions with neighboring galaxies detected also in HI.  By contrast, only four of the twenty-one ($\sim$$19\%$) galaxies in our control sample exhibit HI disturbances.  Removing the one disturbed galaxy since found to be a Seyfert~2 galaxy, only three of the twenty ($15\%$) are actually disturbed in HI.  Only one of the eighteen ($\sim$$6\%$) Seyfert galaxies in the ensemble active galaxy sample exhibit no detectable HI disturbances whatsoever.  By contrast, seventeen of the twenty ($\sim$$85\%$) galaxies in our control sample exhibit no detectable HI disturbances.  These results directly implicate tidal interactions in initiating events that lead to optically-luminous Seyfert activity in a large fraction of local disk galaxies.

\section{SUMMARY AND CONCLUSIONS}
The central purpose of this paper is to determine whether the high incidence of tidal interactions observed in HI gas for a sample of Seyfert galaxies reported in the companion paper by \citet{kuo08} is related to their AGN activity.  Our strategy was to image at the same spatial resolution ($\sim$20~kpc) and sensitivity in HI gas a comparably large number of inactive galaxies that were closely matched in Hubble type and to the degree possible optical luminosity, as well as range in size and inclination, to the active galaxies.  We detected twenty-one of the twenty-seven galaxies in our control sample, imaged at the same spatial resolution and HI column density threshold as the active galaxy sample.  These twenty-one galaxies comprised our ensemble control sample from which we drew the following statistical results.

\begin{itemize}

\item[1.]  Only four of the twenty-one galaxies ($\sim$$19\%$) exhibit spatial and usually also kinematic disturbances on galactic ($\gtrsim 20$~kpc) scales.  One of these disturbed galaxies has since been found to be a Seyfert~2 galaxy, and so only three of the twenty ($15\%$) inactive galaxies in our ensemble control sample are actually disturbed in HI.

\item[2.]  Seventeen of the twenty-one galaxies ($\sim$$81\%$) show no HI disturbances whatsover on galactic ($\gtrsim 20$~kpc) scales.  Excluding again the one disturbed galaxy since found to be a Seyfert~2 galaxy, seventeen of the twenty ($85\%$) inactive galaxies in our ensemble control sample are not disturbed in HI.

\end{itemize}

By contrast, of the eighteen galaxies in the ensemble sample of twenty-three active galaxies classified as Seyferts, seventeen ($\sim$$94\%$) exhibit HI disturbances.  In at least $\sim$$67\%$ and possibly as high as $\sim$$94\%$ of cases, the observed HI disturbances can be traced to tidal interactions with neighboring galaxies detected also in HI.  The dramatic contrast in the incidence of HI disturbances between the active and inactive galaxy samples strongly implicates tidal interactions in initiating events that lead to luminous Seyfert activity in a large fraction of local disk galaxies.

\acknowledgments
\hspace{2.5 cm}
Acknowledgemetns
\newline

Jenny Greene was largely responsible for selecting the control sample. We thank A.-L. Tsai and S.-Y. Lin for reducing part of the data. We thank the important comments from the referee. The National Radio Astronomy Observatory is a facility of the National Science Foundation operated under cooperative agreement by Associated Universities, Inc.  Y.-W. Tang, C.-Y. Kuo, A.-L. Tsai, and S.-Y. Lin all acknowledge the support of a Research Assistantship at the ASIAA where the bulk of this work was done.  J. Lim acknowledges the National Science Council of Taiwan for providing a grant in support of this work.  This research has made use of NASA's Astrophysics Data System Bibliographic Services, and the NASA/IPAC Extragalactic Database (NED) which is operated by the Jet Propulsion Laboratory, California Institute of Technology, under contract with the National Aeronautics and Space Administration.  We acknowledge the usage of the HyperLeda database (http://leda.univ-lyon1.fr).

\begin{figure}
\
\hspace{0.2 cm}
\vspace{-0.1 cm}
\includegraphics[angle=0,scale=0.67]{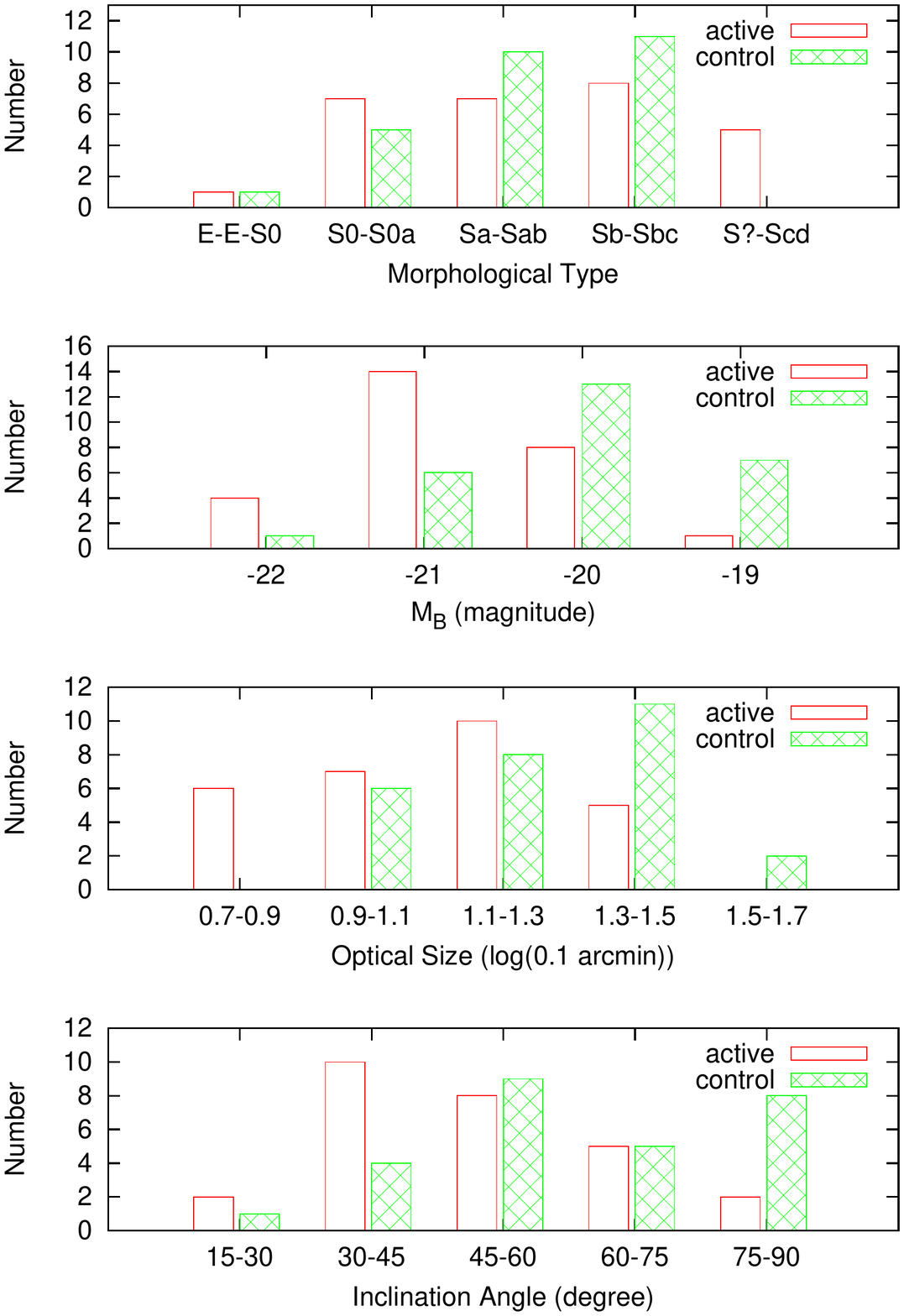}
\vspace{-0.1 cm}
\caption{Number distribution of the active galaxy sample selected
by \citep{kuo08} (unfilled histogram) and the inactive control
sample selected here (hatched histogram) in morphologial type
(upper panel), corrected absolute B-band magnitude (second panel
from top, M$_B$), optical size (second panel from bottom), and
inclination (bottom panel). The morphological type, M$_{B}$,
optical size and inclination angle are retrieved from the
HYPERLEDA database. Note that there are 27 active samples included
in the M$_{B}$ plot. The M$_{B}$ of MS 04595+0327 is not available
in the HYPERLEDA database. The rest of the plots include all the
sample.}
\end{figure}

\begin{figure}
\vspace{-2 cm}
\hspace*{0 cm}
\includegraphics[angle=0, scale=0.75]{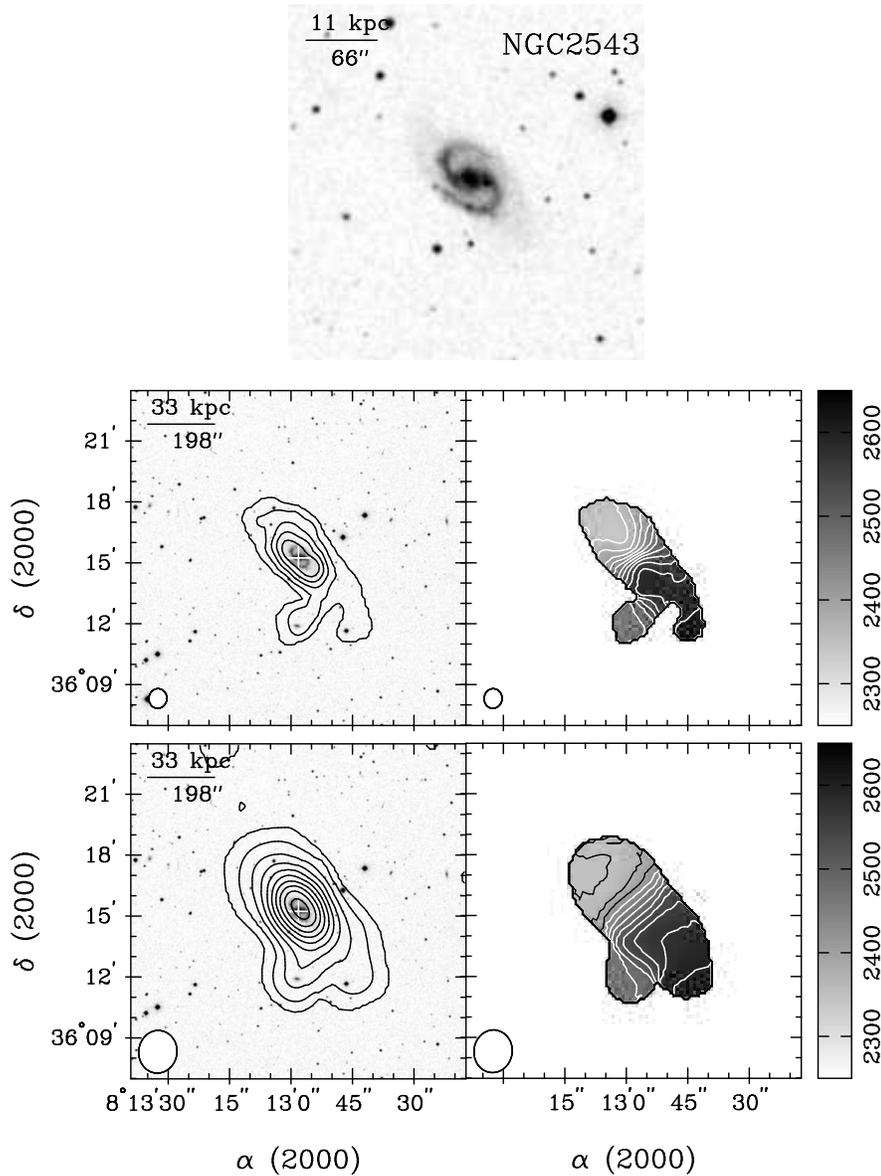}
\vspace{-1.6 cm}
\caption{Upper panel: Optical image of NGC 2543 (Control sample)
from the Second Digitized Sky Survey (DSS2). Middle panels:
(\textsl{Left}) Contours of integrated HI intensity (zeroth
moment) overlaid on the DSS2 image, and (\textsl{Right}) map of
intensity-weighted HI mean velocity (first moment) with full
resolution. Lower panels: (\textsl{Left}) Contours of zeroth
moment overlaid on the DSS2 image and (\textsl{Right}) first
moment map with the same spatial resolution as the Seyfert sample.
In the zeroth moment maps with full resolution, contours are
plotted at 3, 20, 40, 60, $80 \times 20.0 {\rm \ mJy \ beam^{-1}\
km \ s^{-1}} ({\rm 1.0\times10^{19} \ cm^{-2}})$. In the zeroth
moment maps with the same spatial resolution as the Seyfert
sample, contours are plotted at 3, 20, 40, 60, $80 \times 28.3{\rm
\ mJy \ beam^{-1} \ km \ s^{-1}} ({\rm 3.2\times10^{18} \
cm^{-2}})$. In the first moment map, heliocentric velocities are
indicated by the scale wedge (in km/s), and contours plotted at
intervals of $25 {\rm \ km \ s^{-1}}$. The ellipse at the lower
left corner of the lower panels is the half-power width of the
synthesized beam, and has a size of $58\arcsec \times 53 \arcsec$
(full resolution) and $126 \arcsec \times 113\arcsec$ (the same
spatial resolution as the Seyfert sample).}

\end{figure}
\clearpage

\begin{figure}
\vspace{-2.7 cm}
\hspace{-1 cm}
\includegraphics[angle=0, scale=0.8]{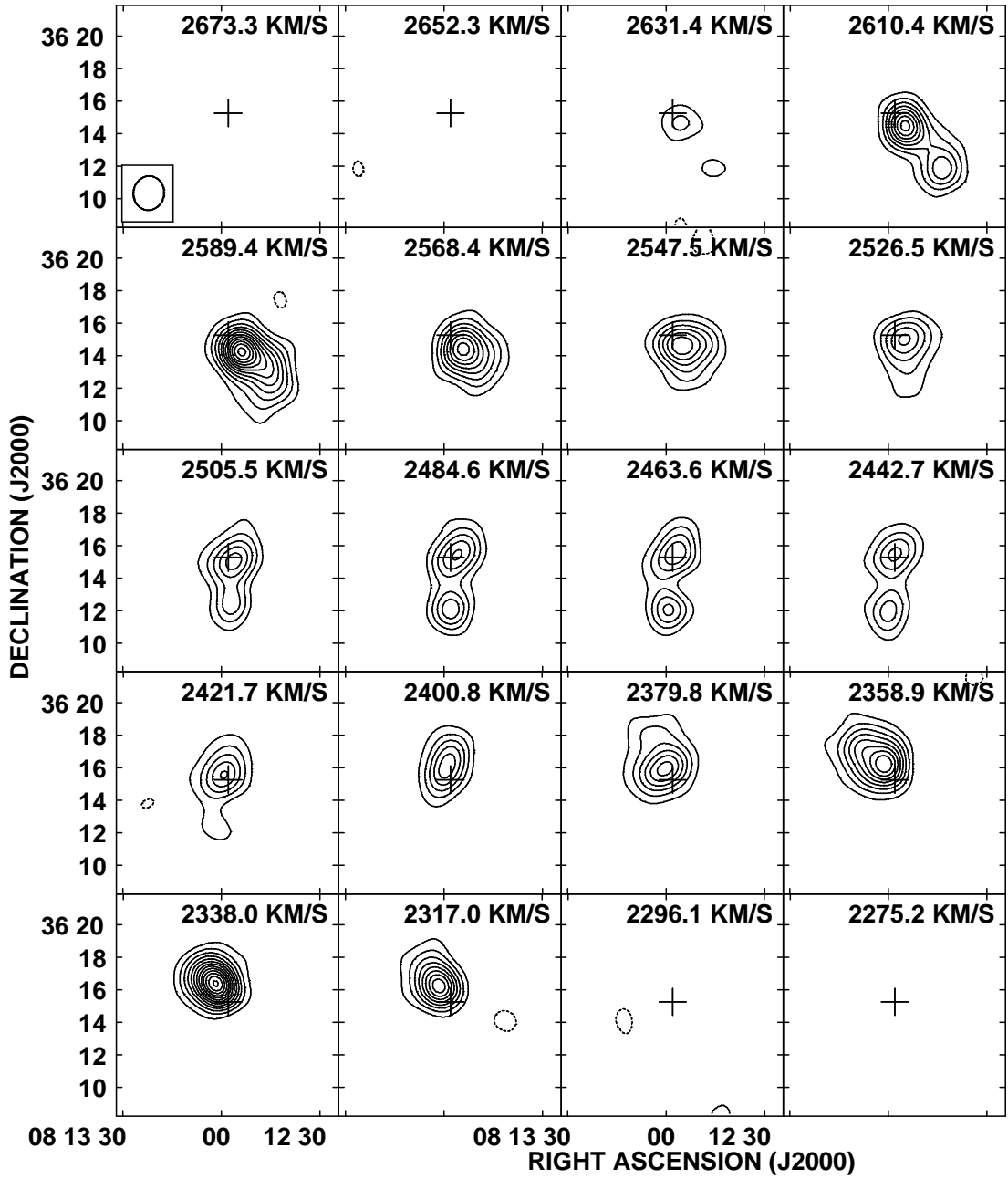}
\vspace{-1.3 cm}
\caption{HI channel maps of NGC 2543 with spatial resolution the
same as the Seyfert sample. Contour levels are plotted at $-3$, 3,
6, 9, 12, 15, $18 \times 0.18 {\rm \ mJy \ beam^{-1}}$ ($1
\sigma$), which corresponds to a HI column density of 2.0 $\times
10^{16} {\rm \ cm^{-2}}$. The central heliocentric velocity is
shown for each channel. The cross marks the position of NGC 2543.
The synthesized beam is shown by the ellipse at the lower left
corner of the top left panel.}

\end{figure}

\clearpage
\begin{figure}
\vspace{-2 cm}
\hspace*{-0.5 cm}
\includegraphics[angle=0, scale=0.8]{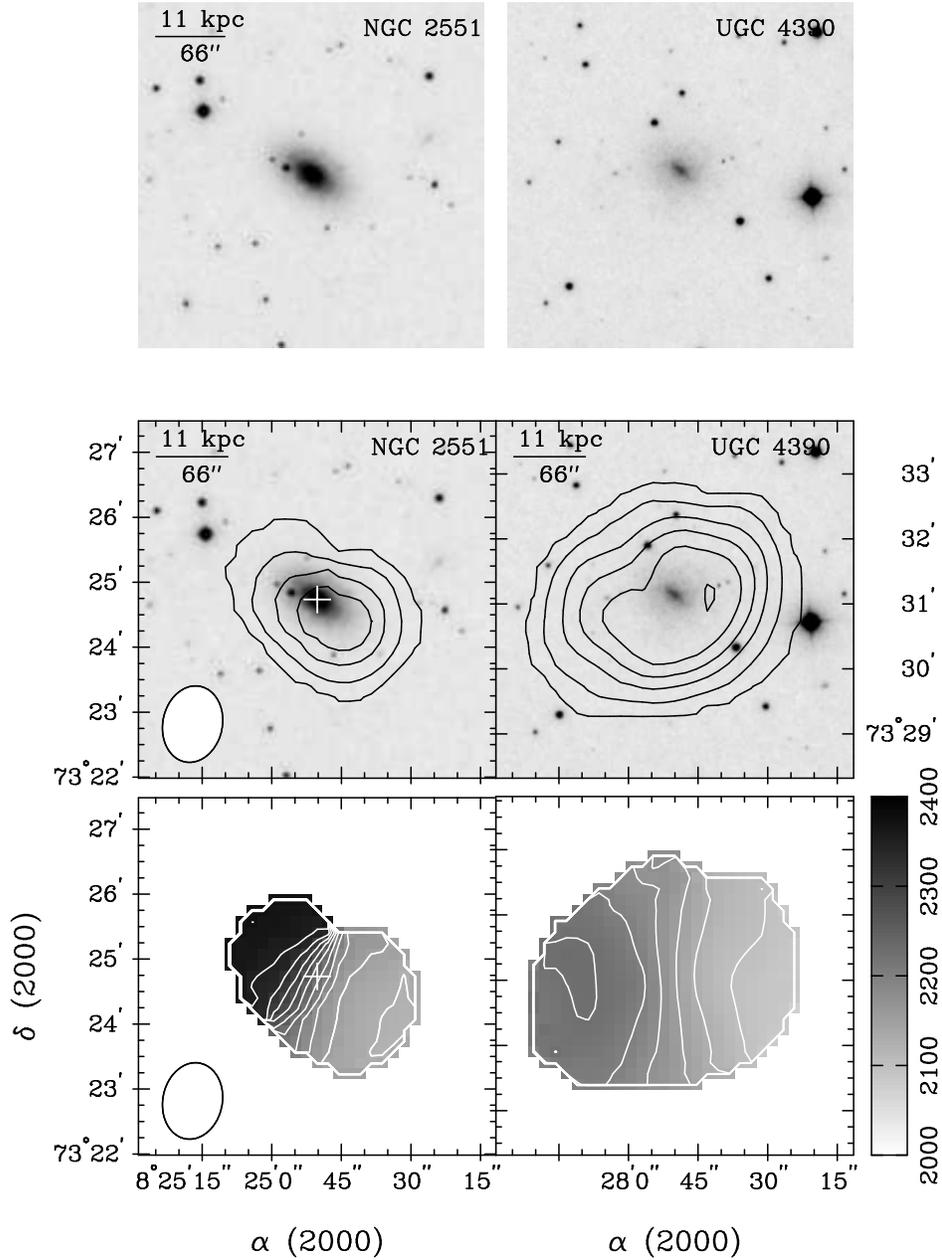}
\vspace{-1.5 cm}
\caption{Upper panels: Optical images of (\textsl{Left}) NGC 2551(Control
sample) and (\textsl{Right}) UGC 4390 from DSS2. Middle panels:
Contours of zeroth moment overlaid on the DSS2 images. Lower
panels: First moment maps. In the zeroth moment map, contours are
plotted at 3, 10, 20, 30, 40
 (NGC 2551) and 3, 6, 9, 12, 15, 18 (UGC 4390) $\times 24.0 {\rm \ mJy
\ beam^{-1} \ km \ s^{-1}} ({\rm 9.8\times10^{18} \ cm^{-2}})$.
The half-power width of the synthesized beam has a size of $72
\arcsec \times 55\arcsec$.}
\end{figure}

\begin{figure}
\vspace{-2 cm}
\hspace*{-0.5 cm}
\includegraphics[angle=0, scale=0.8]{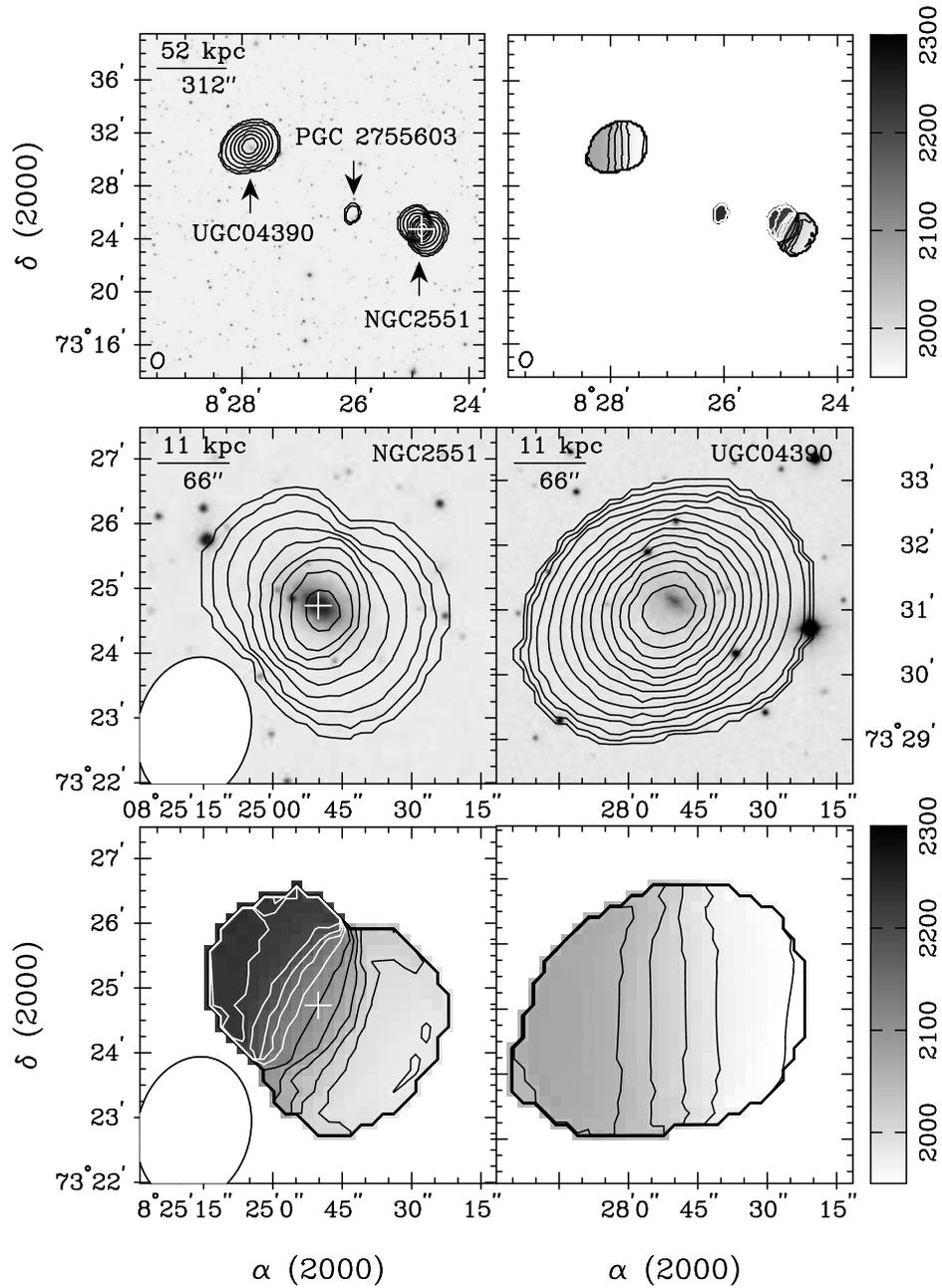}
\vspace{-1.5 cm}
 \caption{Upper panels: (\textsl{Left}) Contours of zeroth moment of NGC 2551 and UGC 4390 overlaid on the DSS2 image and (\textsl{Right}) first moment map with larger field. Middle panels: Contours of zeroth moment with the same spatial resolution
as the Seyfert sample overlaid on the DSS2 images. Lower panels:
First moment maps with the same spatial resolution as the Seyfert
sample. In the zeroth moment maps, contours are plotted at 3, 20,
40, 60, $80 \times 24.0 {\rm \ mJy \ beam^{-1} \ km \ s^{-1}}
({\rm 9.8\times10^{18} \ cm^{-2}})$. The half-power width of the
synthesized beam has a size of $136 \arcsec \times 105\arcsec$.}
\end{figure}

\clearpage
\begin{figure}
\hspace{-1 cm}
\includegraphics[angle=0, scale=0.8]{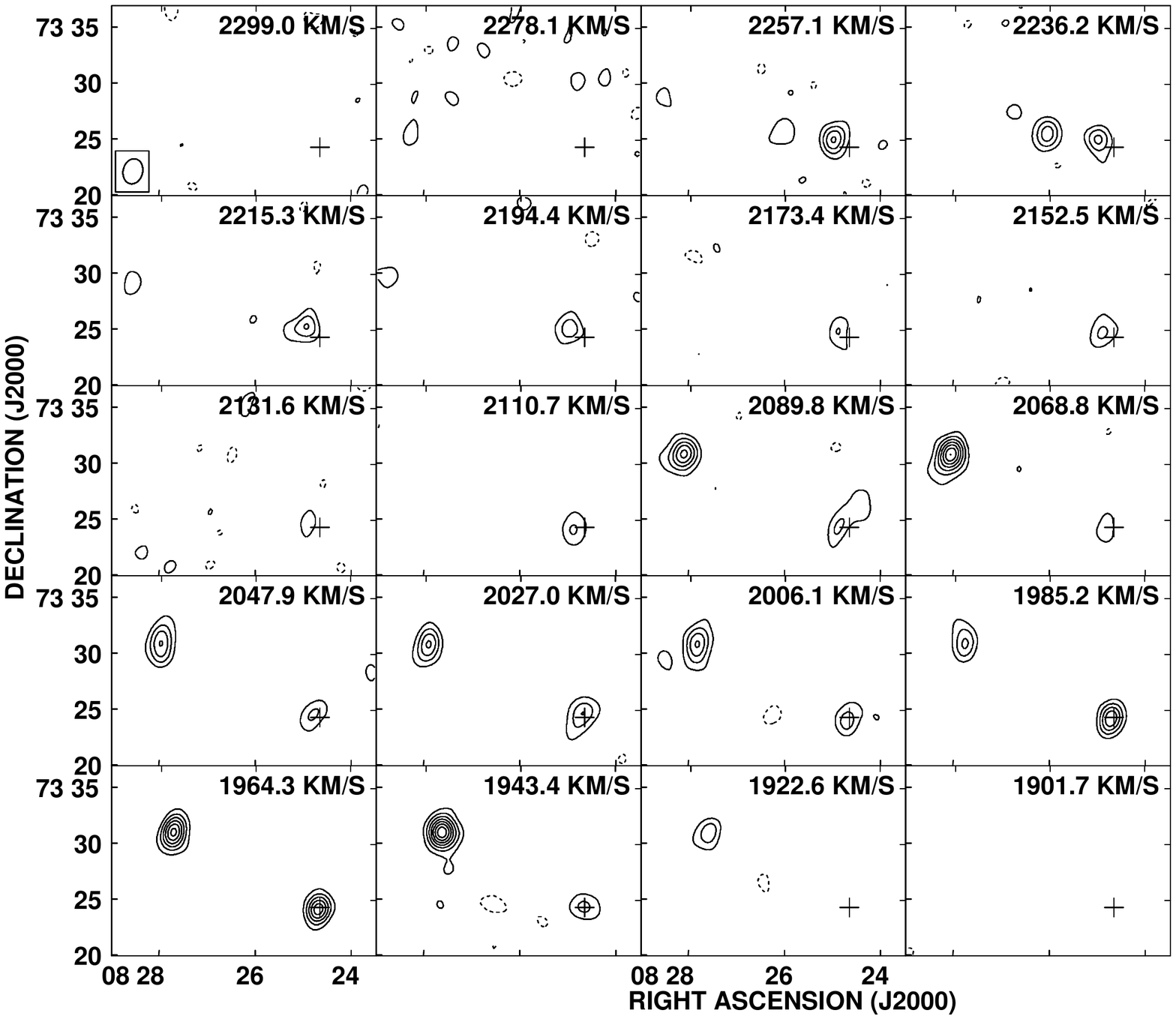}
\caption{HI channel maps of NGC 2551 with spatial resolution the
same as the Seyfert sample. Contour levels are plotted at $-3$, 3,
6, 9, 12, 15, $18 \times 0.18 {\rm \ mJy \ beam^{-1}}$ ($1
\sigma$), which corresponds to a HI column density of 2.0 $\times
10^{16} {\rm \ cm^{-2}}$. The cross marks the position of NGC
2551.}

\end{figure}

\clearpage
\begin{figure}
\vspace{-2 cm}
\hspace*{0.6 cm}
\includegraphics[angle=0, scale=0.7]{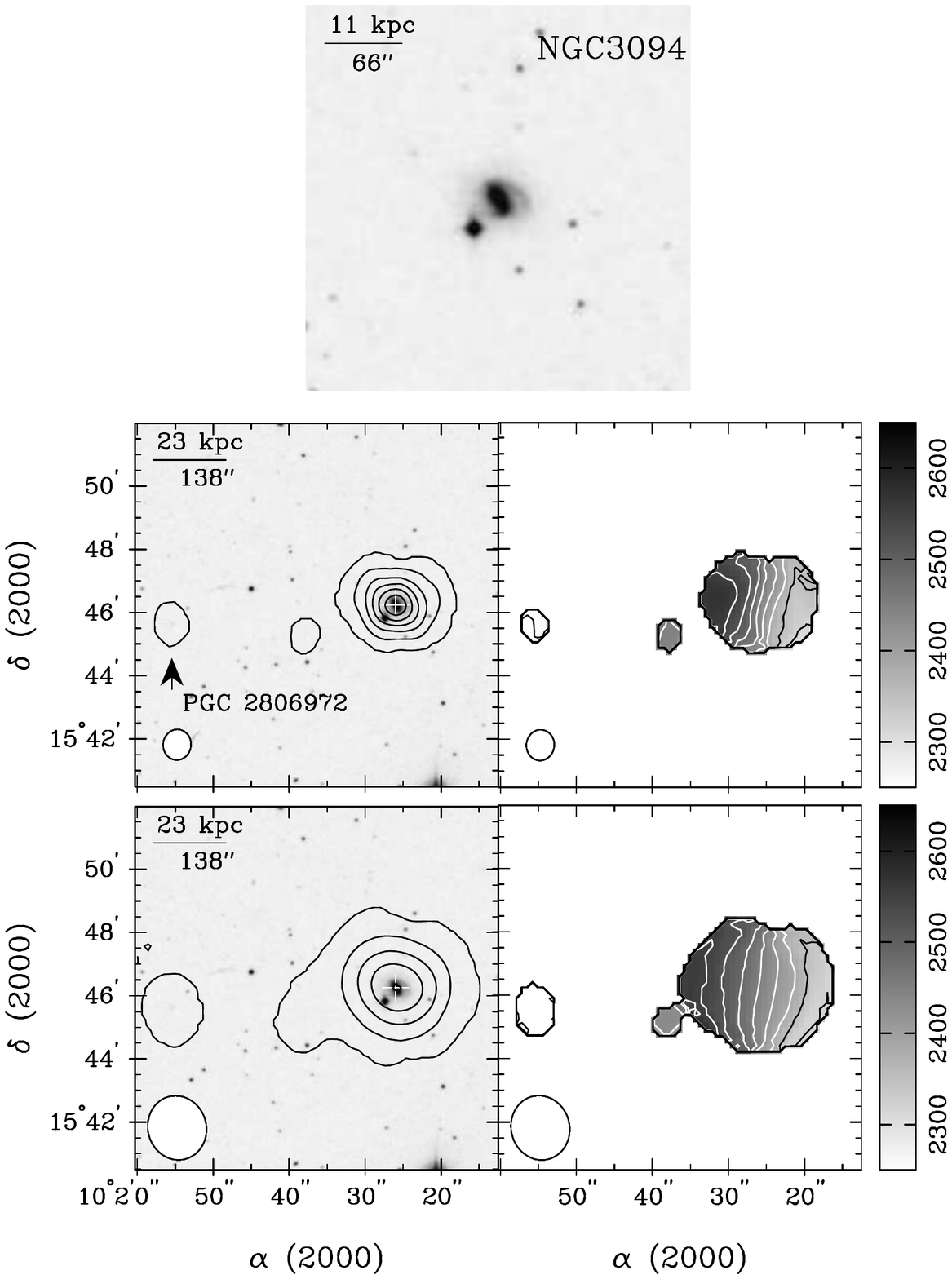}
\vspace{-1.1 cm}
 \caption{Upper panel: Optical images of NGC 3094 (Control sample) from DSS2.
Middle panels: (\textsl{Left}) Contours of zeroth moment overlaid
on the DSS2 image and (\textsl{Right}) first moment map with full
resolution. Lower panels:(\textsl{Left}) Contours of zeroth moment
and (\textsl{Right}) first moment map with the same spatial
resolution as the Seyfert sample. In the zeroth moment maps with
full resolution, contours are plotted at 3, 20, 40, 60, $80 \times
28.2{\rm \ mJy \ beam^{-1} \ km \ s^{-1}} ({\rm 1.5\times10^{19} \
cm^{-2}})$. In the zeroth moment maps with the same spatial
resolution as the Seyfert sample, contours are plotted at 3, 20,
40, 60, $80 \times 71.7 {\rm \ mJy \ beam^{-1} \ km \ s^{-1}}
({\rm 8.5\times10^{18} \ cm^{-2}})$. The half-power width of the
synthesized beam has a size of $59\arcsec \times 52\arcsec$ (full
resolution) and $123 \arcsec \times 111\arcsec$ (the same spatial
resolution as the Seyfert sample).}
\end{figure}

\clearpage
\begin{figure}
\hspace{-2 cm}
\includegraphics[angle=0, scale=0.8]{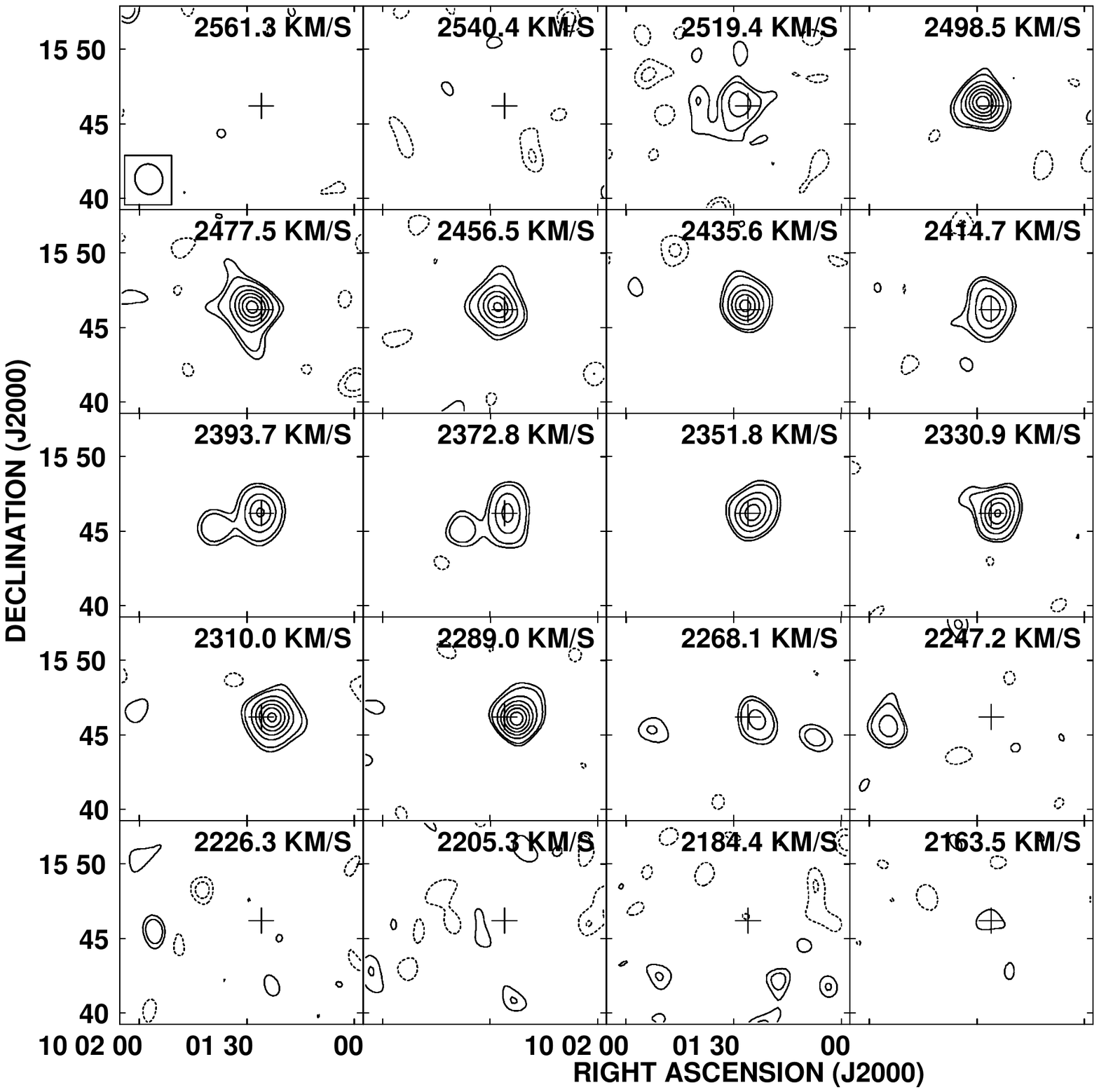}
\caption{HI channel maps of NGC 3094 with spatial resolution the
same as the Seyfert sample. Contour levels are plotted at $-3$,
$-2$, 2, 3, 6, 9, 12, $15 \times 0.18 {\rm \ mJy \ beam^{-1}}$ ($1
\sigma$), which corresponds to a HI column density of 2.0 $\times
10^{16} {\rm \ cm^{-2}}$. The cross marks the position of NGC
3094.}

\end{figure}

\clearpage
\begin{figure}
\vspace{-2 cm}
\hspace*{0.6 cm}
\includegraphics[angle=0, scale=0.7]{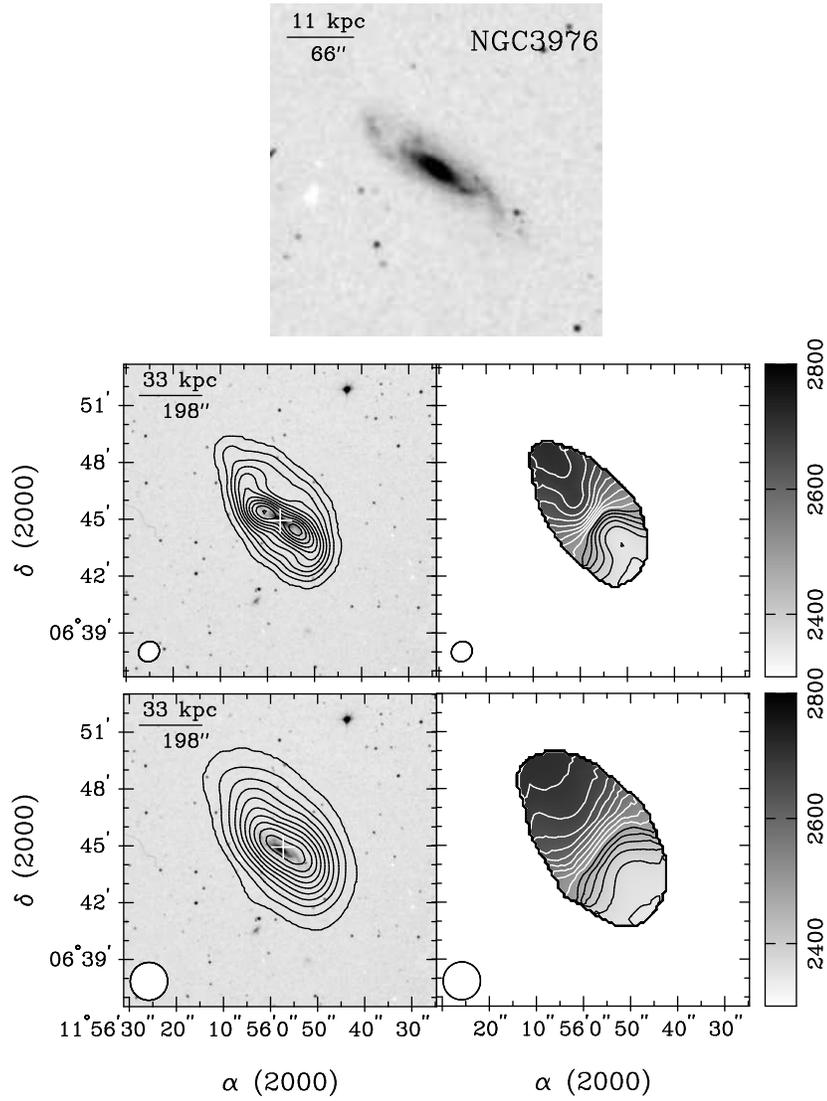}
\vspace{-1.1 cm}
\caption{Upper panel: Optical images of NGC 3976 (Control sample)
from DSS2. Middle panels: (\textsl{Left}) Contours of zeroth
moment overlaid on the DSS2 image and (\textsl{Right}) first
moment map with full resolution. Lower panels: (\textsl{Left})
Contours of zeroth moment and (\textsl{Right}) first moment map
with the same spatial resolution as the Seyfert sample. In the
zeroth moment maps with full resolution, contours are plotted at
3, 20, 40, 60, $80 \times 28.2 {\rm \ mJy \ beam^{-1} \ km \
s^{-1}} ({\rm 1.1\times10^{19} \ cm^{-2}})$. In the zeroth moment
maps with the same spatial resolution as the Seyfert sample,
contours are plotted at 3, 20, 40, 60, $80 \times 71.2 {\rm \ mJy
\ beam^{-1} \ km \ s^{-1}} ({\rm 8.1\times10^{18} \ cm^{-2}})$.
The half-power width of the synthesized beam has a size of
$68\arcsec \times 62\arcsec$ (full resolution) and $120 \arcsec
\times 118\arcsec$ (the same spatial resolution as the Seyfert
sample).}
\end{figure}

\clearpage
\begin{figure}
\vspace{-2 cm}
\hspace*{-0.5 cm}
\includegraphics[angle=0, scale=0.8]{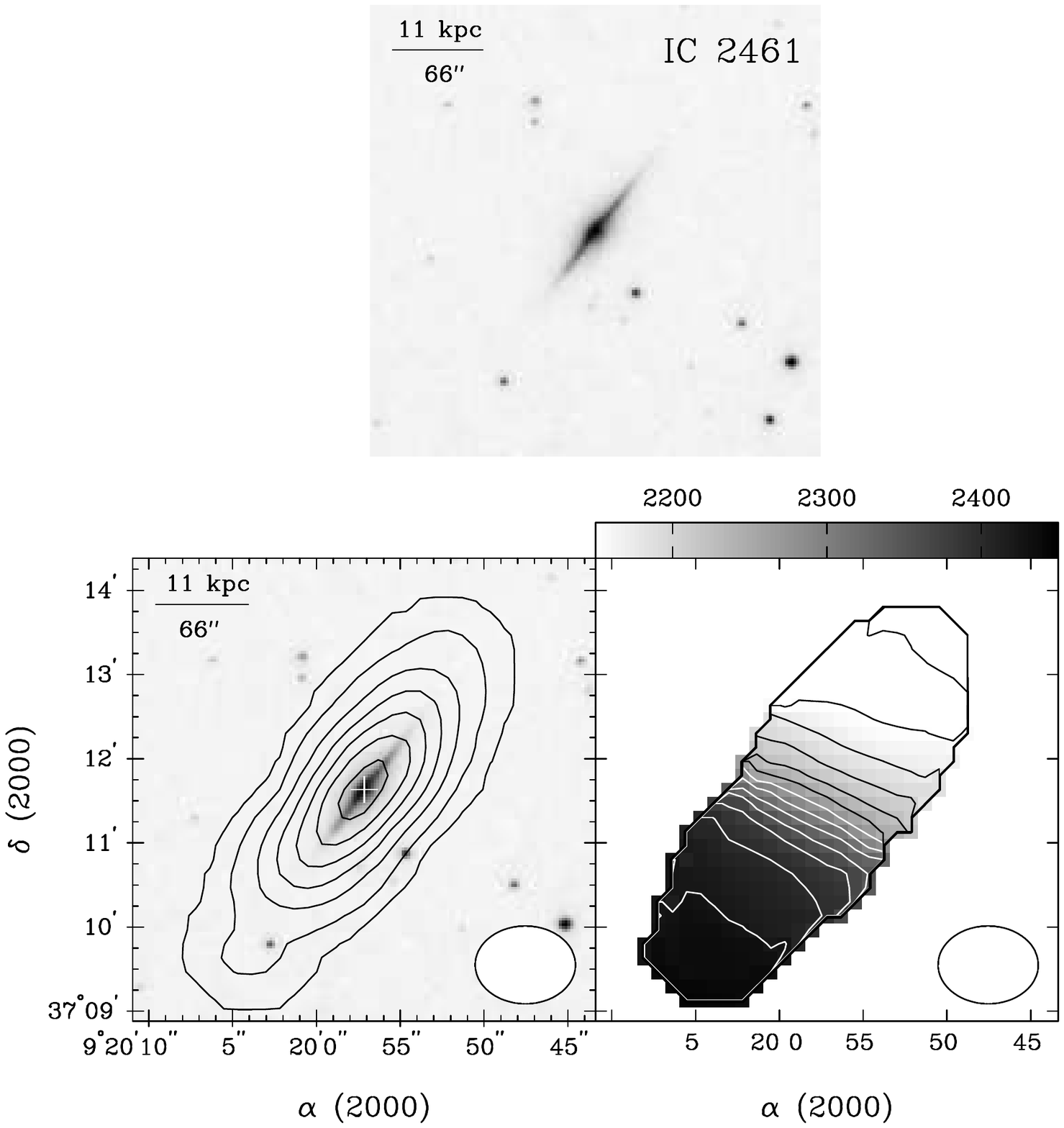}
\vspace{-2.5 cm}
\caption{Upper panel: Optical image of IC 2461 (Control sample)
from the DSS2.  Lower panels: (\textsl{Left}) Contours of zeroth
moment overlaid on the DSS2 image, and (\textsl{Right}) first
moment map. In the zeroth moment map, contours are plotted at 3,
20, 40, 60, $80 \times 26.8 {\rm \ mJy \ beam^{-1}  \ km \ s^{-1}}
(1.1 \times 10^{19} {\rm \ cm^{-2}})$. The half-power width of the
synthesized beam has a size of $72 \arcsec \times 55\arcsec$.}

\label{fig1a}
\end{figure}

\clearpage
\begin{figure}
\vspace{-2 cm}
\hspace*{-0.5 cm}
\includegraphics[angle=0, scale=0.8]{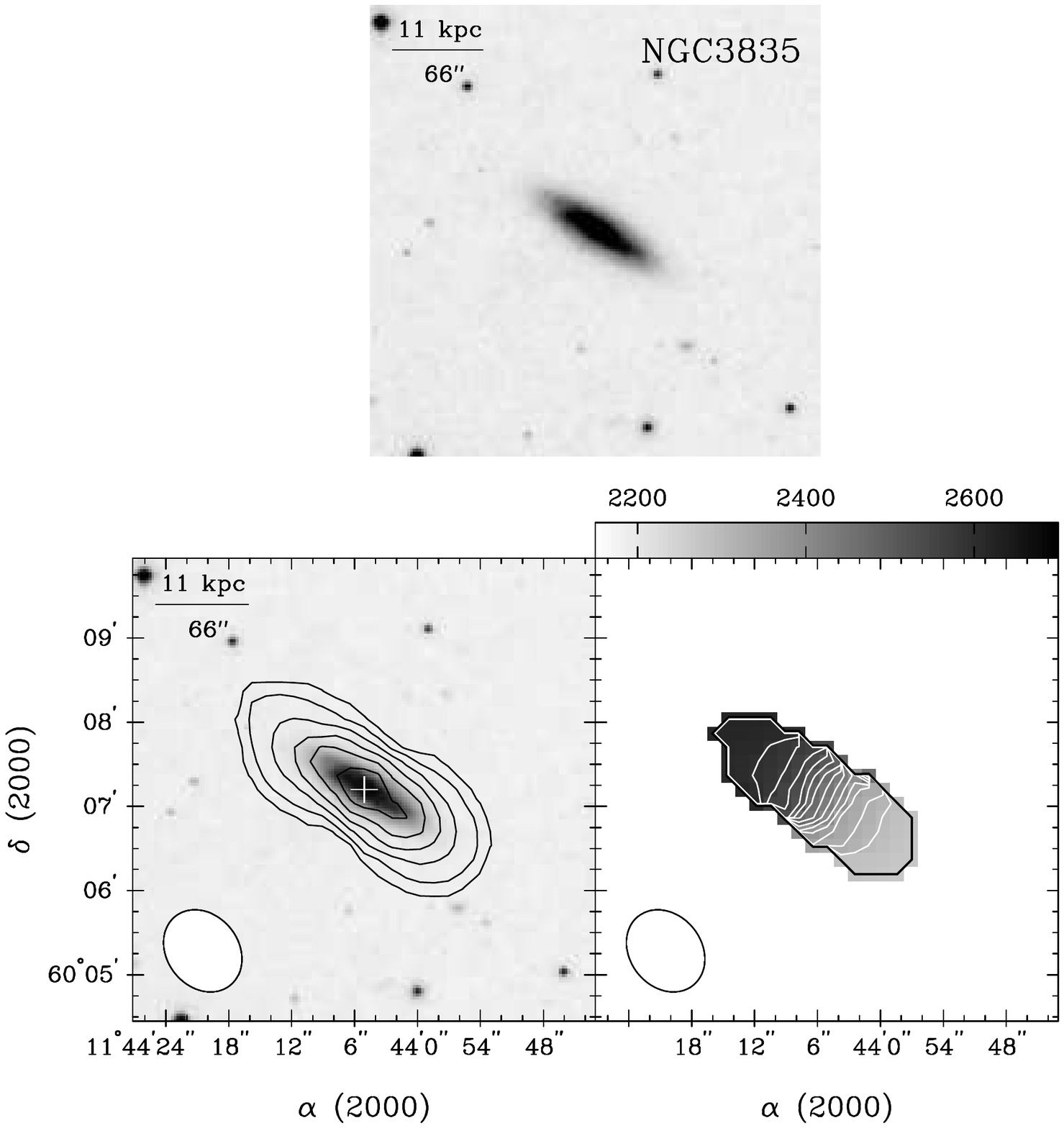}
\vspace{-2.5 cm}
\caption{Upper panel: Optical image of NGC 3835 (Control sample)
from the DSS2. Lower panels: (\textsl{Left}) Contours of zeroth
moment overlaid on the DSS2 image, and (\textsl{Right}) first
moment map. In the zeroth moment map, contours are plotted at 3,
10, 20, 30, $40 \times 31.2 {\rm \ mJy \ beam^{-1} \ km \ s^{-1}}$
($1.6 \times 10^{19} {\rm \ cm^{-2}}$). The half-power width of
the synthesized beam has a size of $63 \arcsec \times 51\arcsec$.}

\end{figure}

\clearpage
\begin{figure}
\vspace{-2 cm}
\hspace*{-0.5 cm}
\includegraphics[angle=0, scale=0.8]{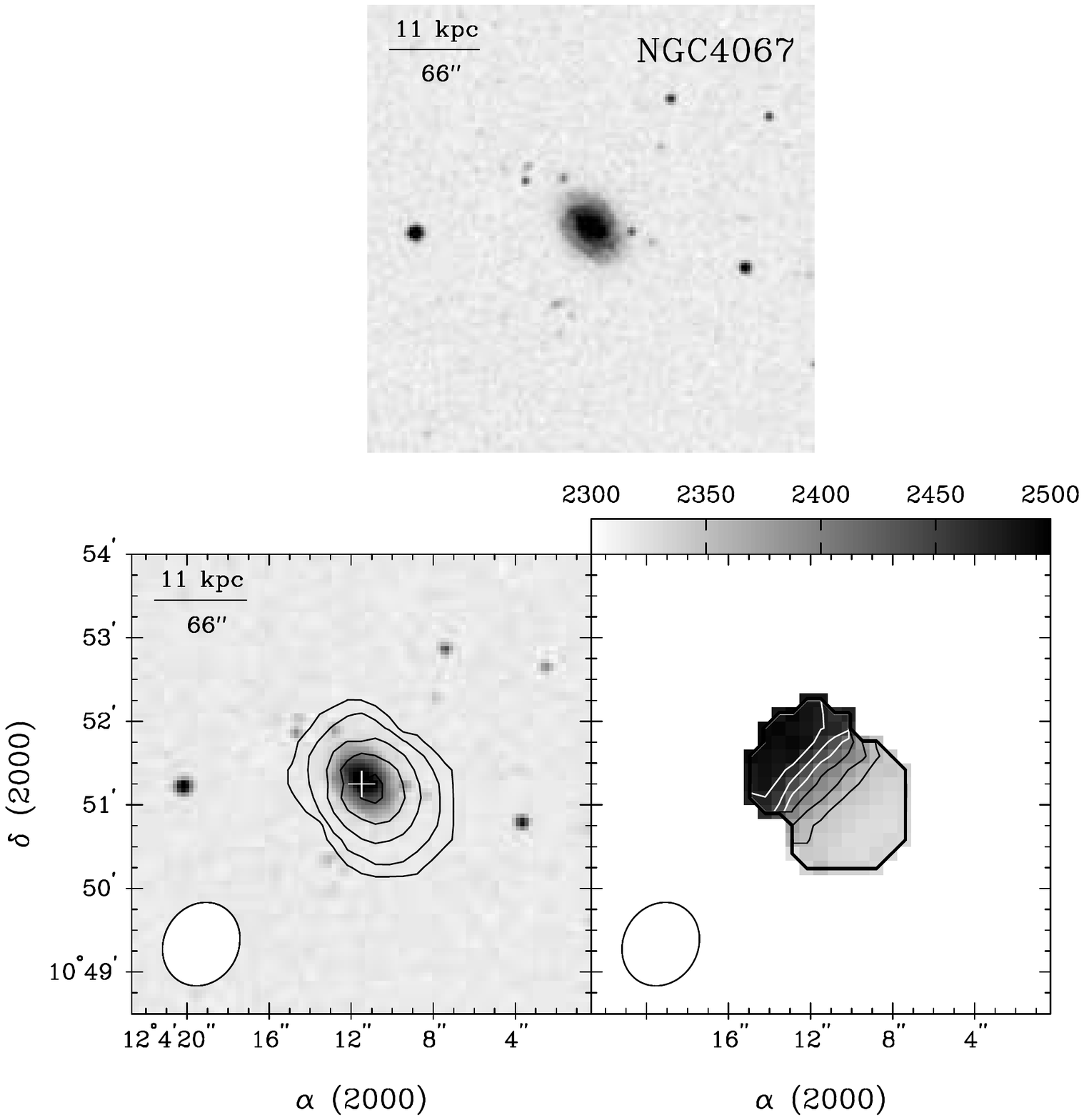}
\vspace{-2.5 cm}
\caption{Upper panel: Optical image of NGC 4067 (Control sample)
from the DSS2. Lower panels: (\textsl{Left}) Contours of zeroth
moment overlaid on the DSS2 image, and (\textsl{Right}) first
moment map. In the zeroth moment map, contours are plotted at 3,
10, 20, 30, $40 \times 31.2 {\rm \ mJy \ beam^{-1} \ km \ s^{-1}}$
($1.5 \times 10^{19} {\rm \ cm^{-2}}$). The half-power width of
the synthesized beam has a size of $62 \arcsec \times 54\arcsec$.}
\end{figure}

\clearpage
\begin{figure}
\vspace{-2 cm}
\hspace*{-0.5 cm}
\includegraphics[angle=0, scale=0.8]{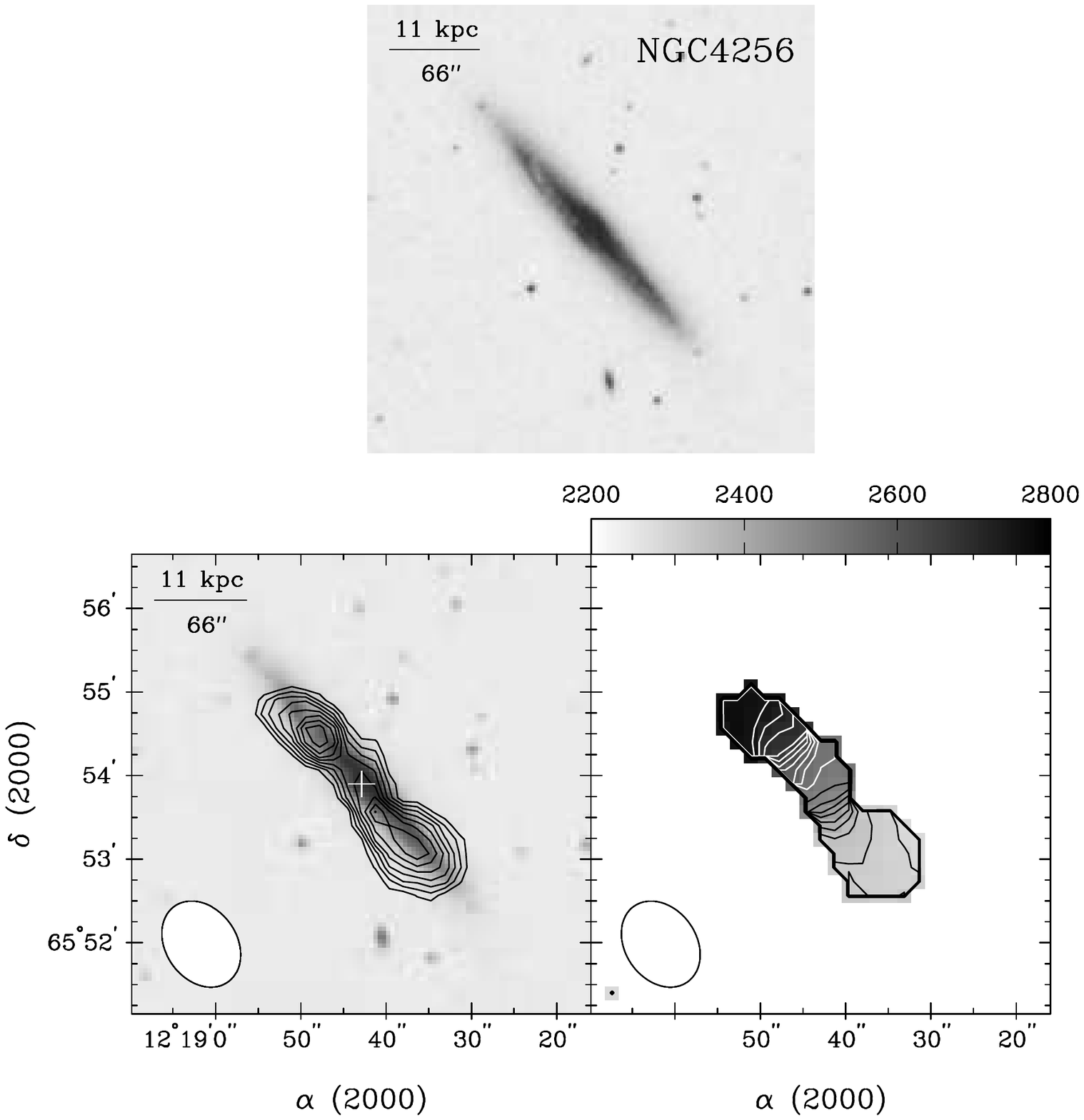}
\vspace{-2.5 cm}
\caption{Upper panel: Optical image of NGC 4256 (Control sample)
from the DSS2. Lower panels: (\textsl{Left}) Contours of zeroth
moment overlaid on the DSS2 image, and (\textsl{Right}) first
moment map. In the zeroth moment map, contours are plotted at 3,
6, 9, 12, $15 \times 38.0 {\rm \ mJy \ beam^{-1} \ km \ s^{-1}}$
($1.8 \times 10^{19} {\rm \ cm^{-2}}$). The half-power width of
the synthesized beam has a size of $67 \arcsec \times 51\arcsec$.}

\end{figure}

\clearpage
\begin{figure}
\vspace{-2 cm}
\hspace*{-0.5 cm}
\includegraphics[angle=0, scale=0.8]{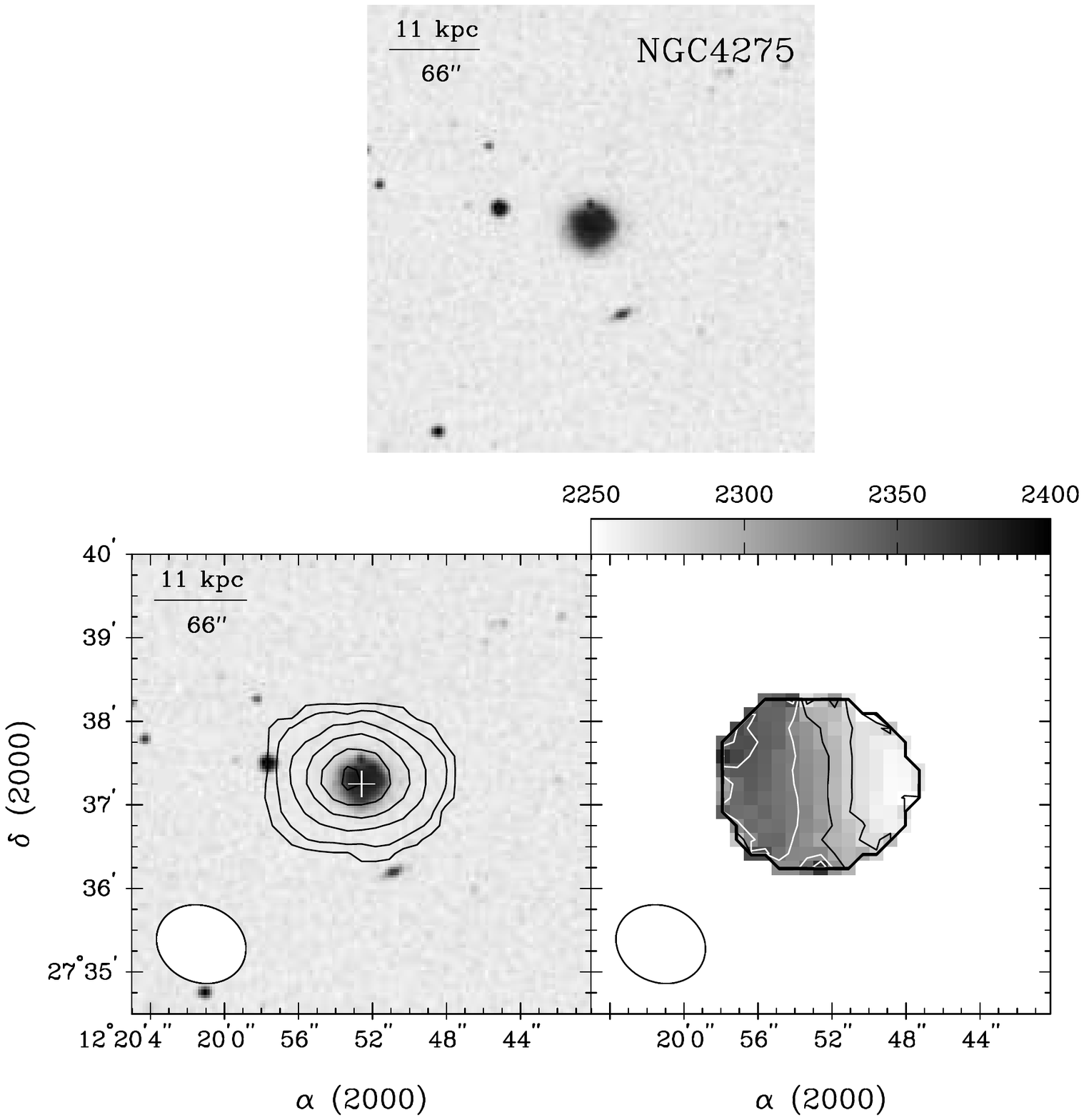}
\vspace{-2.5 cm}
\caption{Upper panel: Optical image of NGC 4275 (Control sample)
from the DSS2. Lower panels: (\textsl{Left}) Contours of zeroth
moment overlaid on the DSS2 image, and (\textsl{Right}) first
moment map. In the zeroth moment map, contours are plotted at 3,
10, 20, 30, $40 \times 23.2 {\rm \ mJy \ beam^{-1} \ km \ s^{-1}}$
($1.1 \times 10^{19} {\rm \ cm^{-2}}$). The half-power width of
the synthesized beam has a size of $63 \arcsec \times 55\arcsec$.}
\end{figure}

\clearpage
\begin{figure}
\vspace{-2 cm}
\hspace*{-0.5 cm}
\includegraphics[angle=0, scale=0.8]{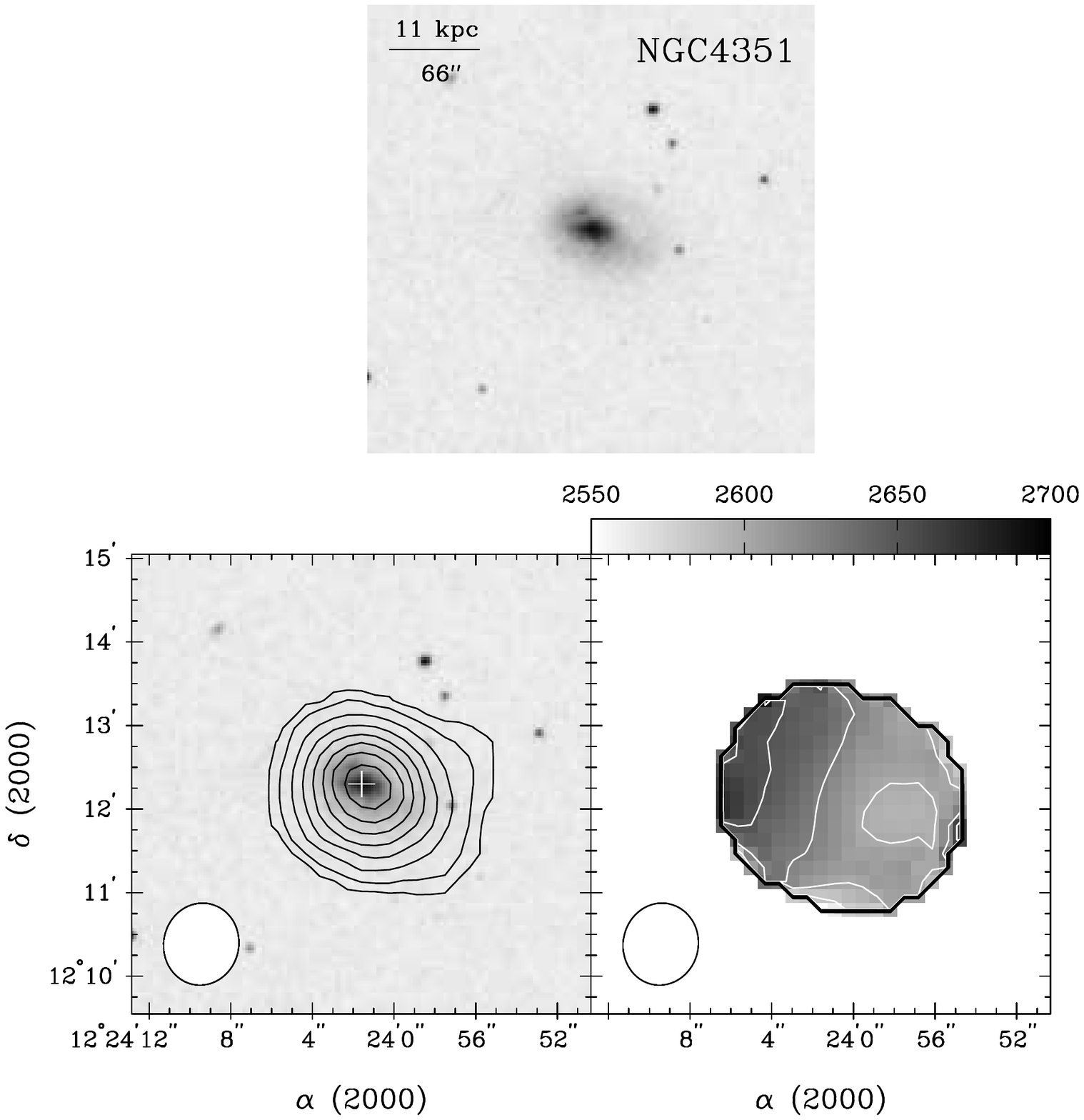}
\vspace{-2.5 cm}
\caption{Upper panel: Optical image of NGC 4351 (Control sample)
from the DSS2. Lower panels: (\textsl{Left}) Contours of zeroth
moment overlaid on the DSS2 image, and (\textsl{Right}) first
moment map. In the zeroth moment map, contours are plotted at 3,
10, 20, 30, $40 \times 20.8 {\rm \ mJy \ beam^{-1} \ km \ s^{-1}}$
($1.1 \times 10^{19} {\rm \ cm^{-2}}$). The half-power width of
the synthesized beam has a size of $59 \arcsec \times 54\arcsec$.}

\end{figure}

\clearpage
\begin{figure}
\vspace{-2 cm}
\hspace*{-0.5 cm}
\includegraphics[angle=0, scale=0.8]{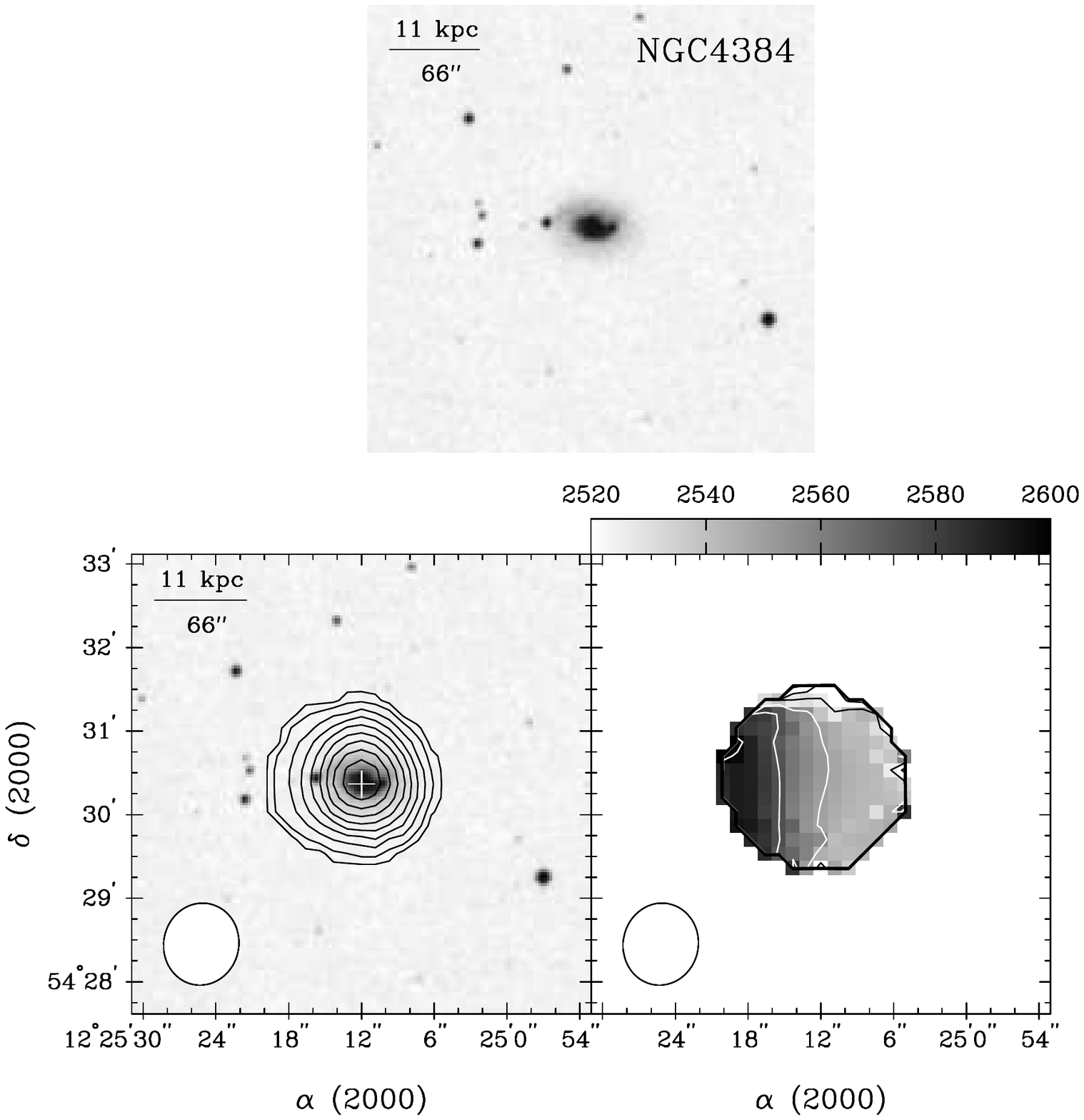}
\vspace{-2.5 cm}
\caption{Upper panel: Optical image of NGC 4384 (Control sample)
from the DSS2. Lower panels: (\textsl{Left}) Contours of zeroth
moment overlaid on the DSS2 image, and (\textsl{Right}) first
moment map. In the zeroth moment map, contours are plotted at 3,
10, 20, 30, $40 \times 18.0 {\rm \ mJy \ beam^{-1} \ km \ s^{-1}}$
($7.8 \times 10^{18} {\rm \ cm^{-2}}$). The half-power width of
the synthesized beam has a size of $66 \arcsec \times 56\arcsec$.}

\end{figure}

\clearpage
\begin{figure}
\vspace{-2 cm}
\hspace*{-0.5 cm}
\includegraphics[angle=0, scale=0.8]{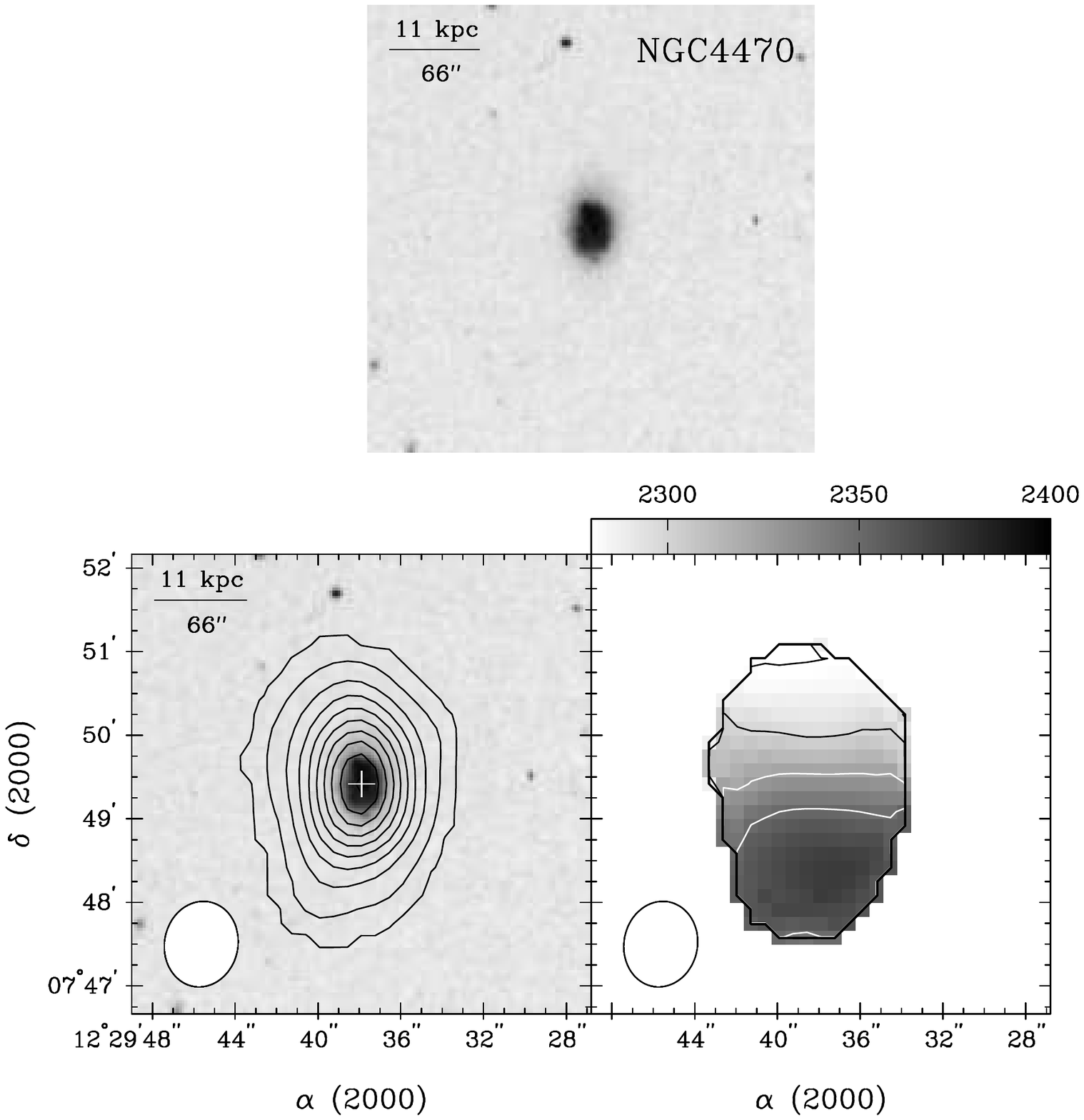}
\vspace{-2.5 cm}
\caption{Upper panel: Optical image of NGC 4470 (Control sample)
from the DSS2. Lower panels: (\textsl{Left}) Contours of zeroth
moment overlaid on the DSS2 image, and (\textsl{Right}) first
moment map. In the zeroth moment map, contours are plotted at 3,
20, 40, 60, $80 \times 17.0 {\rm \ mJy \ beam^{-1} \ km \ s^{-1}}$
($8.4 \times 10^{18} {\rm \ cm^{-2}}$). The half-power width of
the synthesized beam has a size of $62 \arcsec \times 53\arcsec$.}

\end{figure}

\clearpage
\begin{figure}
\vspace{-2 cm}
\hspace*{-0.5 cm}
\includegraphics[angle=0, scale=0.8]{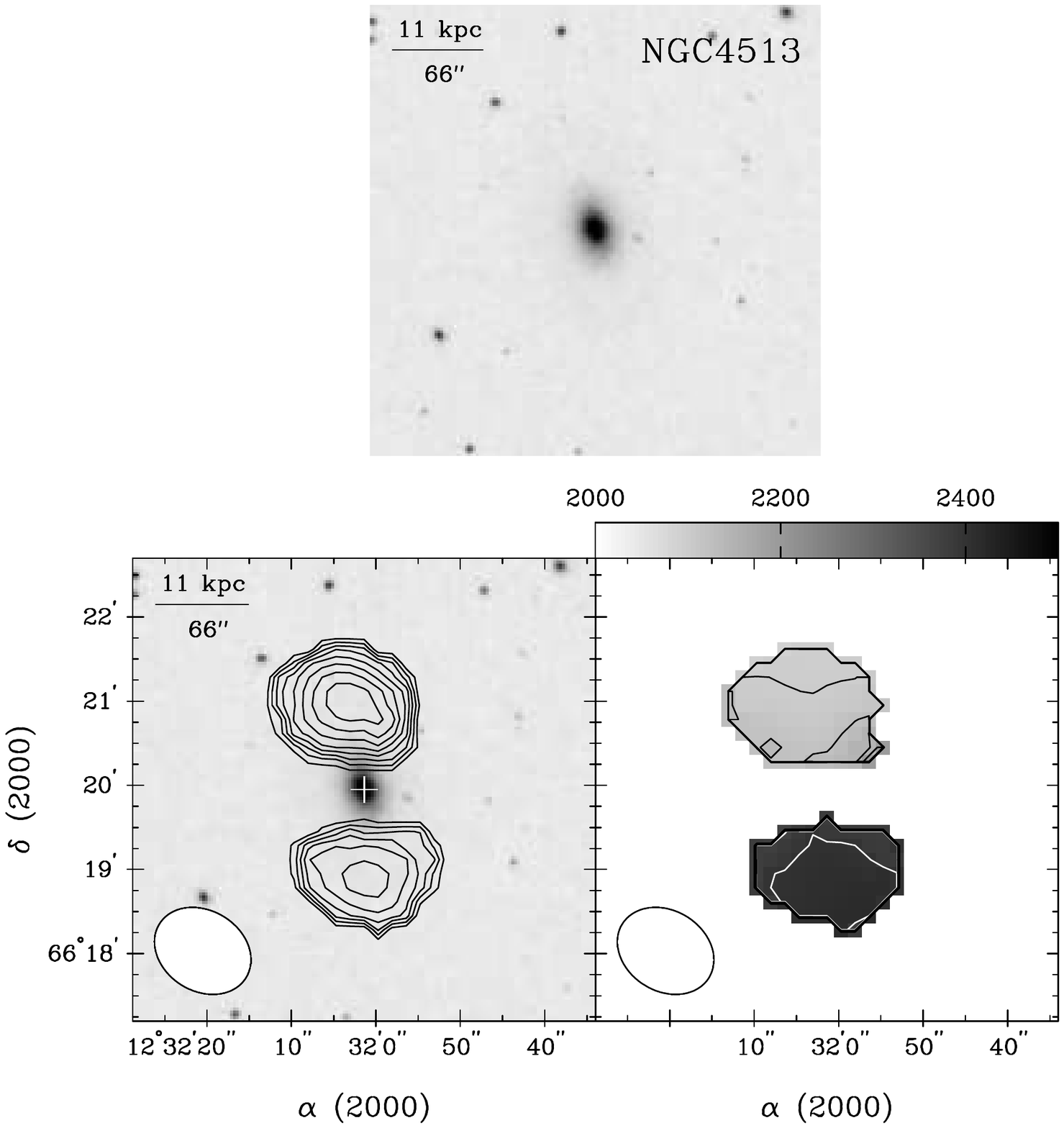}
\vspace{-2.5 cm}
\caption{Upper panel: Optical image of NGC 4513 (Control sample)
from the DSS2. Lower panels: (\textsl{Left}) Contours of zeroth
moment overlaid on the DSS2 image, and (\textsl{Right}) first
moment map. In the zeroth moment map, contours are plotted at 3,
6, 9, 12, $15\times 31.2 {\rm \ mJy \ beam^{-1} \ km \ s^{-1}}$
($1.2 \times 10^{19} {\rm \ cm^{-2}}$). The half-power width of
the synthesized beam has a size of $73 \arcsec \times 57\arcsec$.}

\end{figure}

\clearpage
\begin{figure}
\vspace{-2 cm}
\hspace*{-0.5 cm}
\includegraphics[angle=0, scale=0.8]{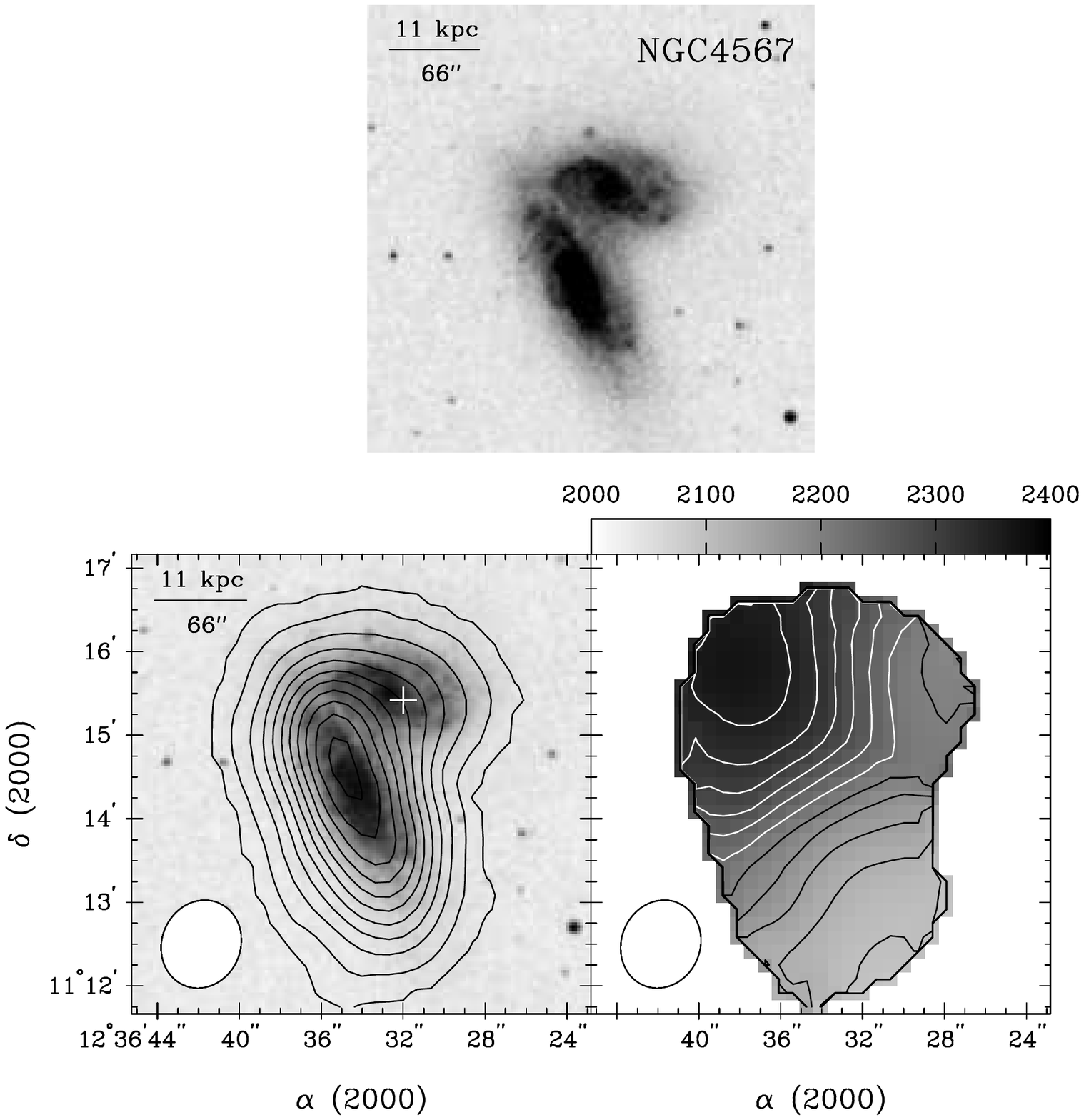}
\vspace{-2.5 cm}
\caption{Upper panel: Optical image of NGC 4567 (Control sample)
from the DSS2. Lower panels: (\textsl{Left}) Contours of zeroth
moment overlaid on the DSS2 image, and (\textsl{Right}) first
moment map. In the zeroth moment map, contours are plotted at 3,
20, 40, 60, $80 \times 24.8 {\rm \ mJy \ beam^{-1} \ km \ s^{-1}}$
($1.1 \times 10^{19} {\rm \ cm^{-2}}$). The half-power width of
the synthesized beam has a size of $64 \arcsec \times 57\arcsec$.}

\end{figure}

\clearpage
\begin{figure}
\vspace{-2 cm}
\hspace*{-0.5 cm}
\includegraphics[angle=0, scale=0.8]{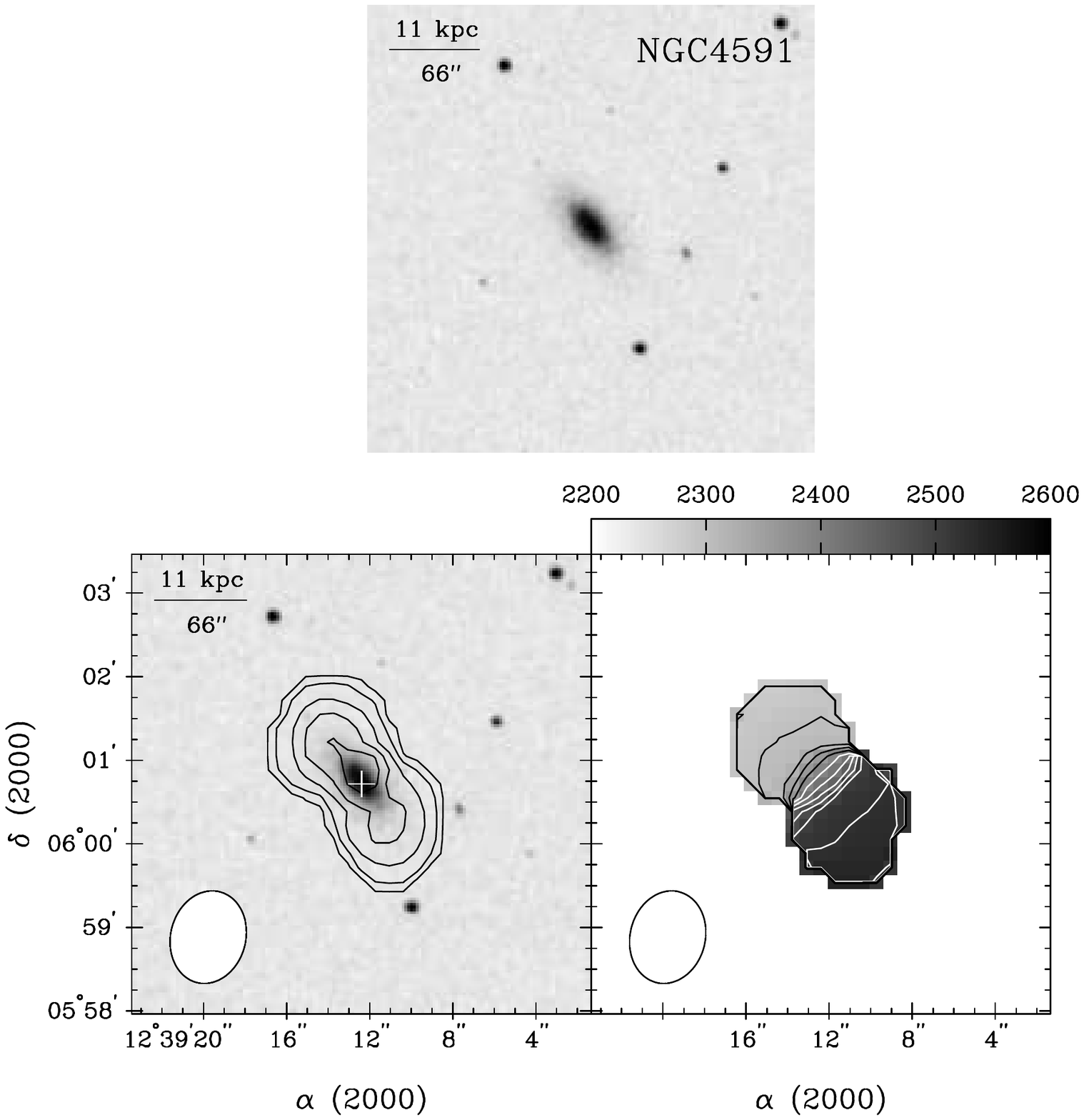}
\vspace{-2.5 cm}
\caption{Upper panel: Optical image of NGC 4591 (Control sample)
from the DSS2. Lower panels: (\textsl{Left}) Contours of zeroth
moment overlaid on the DSS2 image, and (\textsl{Right}) first
moment map. In the zeroth moment map, contours are plotted at 3,
10, 20, 30, $40 \times 24.0 {\rm \ mJy \ beam^{-1} \ km \ s^{-1}}$
($1.1 \times 10^{19} {\rm \ cm^{-2}}$). The half-power width of
the synthesized beam has a size of $67 \arcsec \times 54\arcsec$.}

\end{figure}

\clearpage
\begin{figure}
\vspace{-2 cm}
\hspace*{-0.5 cm}
\includegraphics[angle=0, scale=0.8]{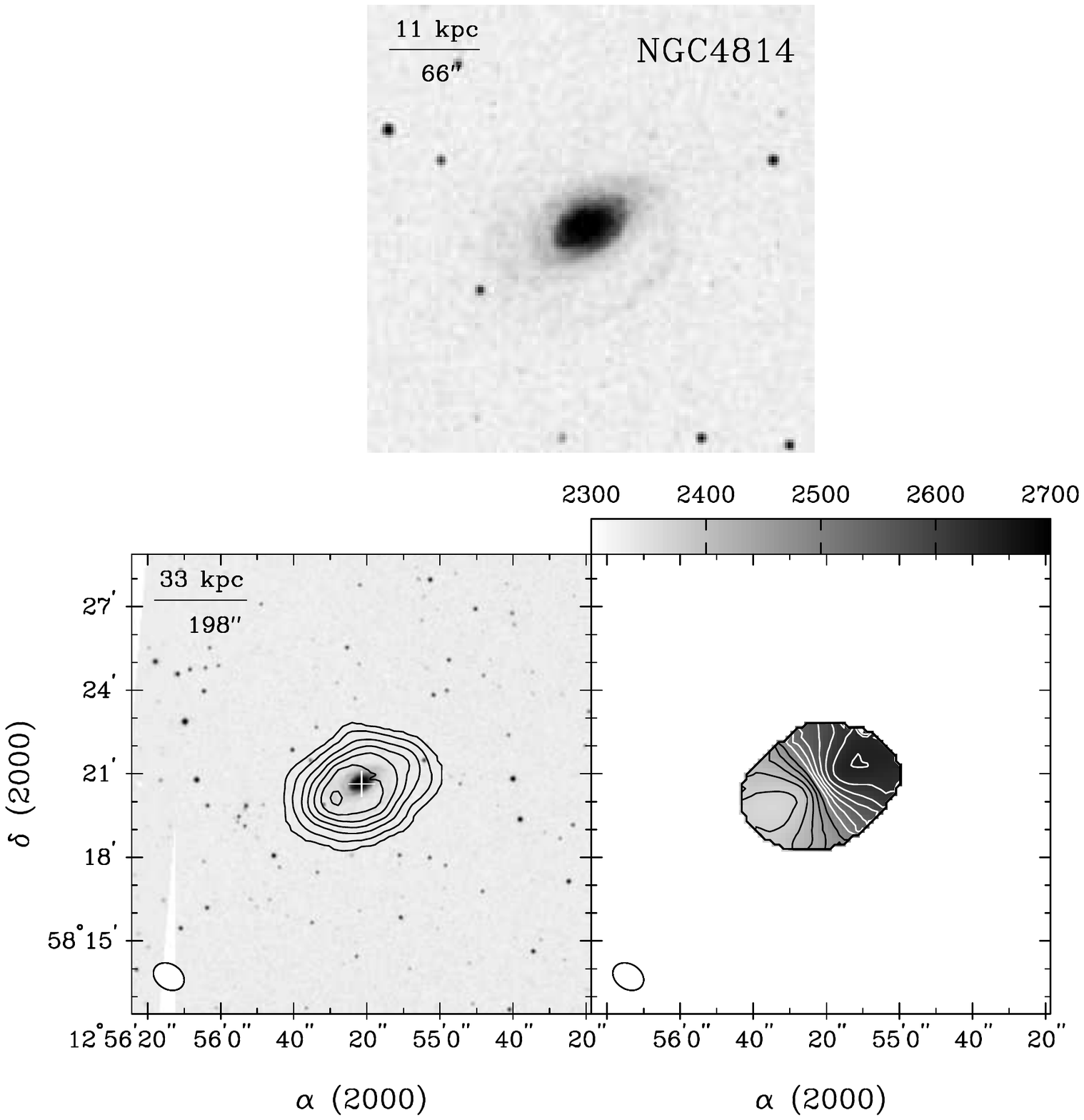}
\vspace{-2.5 cm}
\caption{Upper panel: Optical image of NGC 4814 (Control sample)
from the DSS2. Lower panels: (\textsl{Left}) Contours of zeroth
moment overlaid on the DSS2 image, and (\textsl{Right}) first
moment map. In the zeroth moment map, contours are plotted at 3,
10, 20, 30, $40 \times 46.1 {\rm \ mJy \ beam^{-1} \ km \ s^{-1}}$
($2.0 \times 10^{19} {\rm \ cm^{-2}}$). The half-power width of
the synthesized beam has a size of $71 \arcsec \times 54\arcsec$.}

\end{figure}

\clearpage
\begin{figure}
\vspace{-2 cm}
\hspace*{-0.5 cm}
\includegraphics[angle=0, scale=0.8]{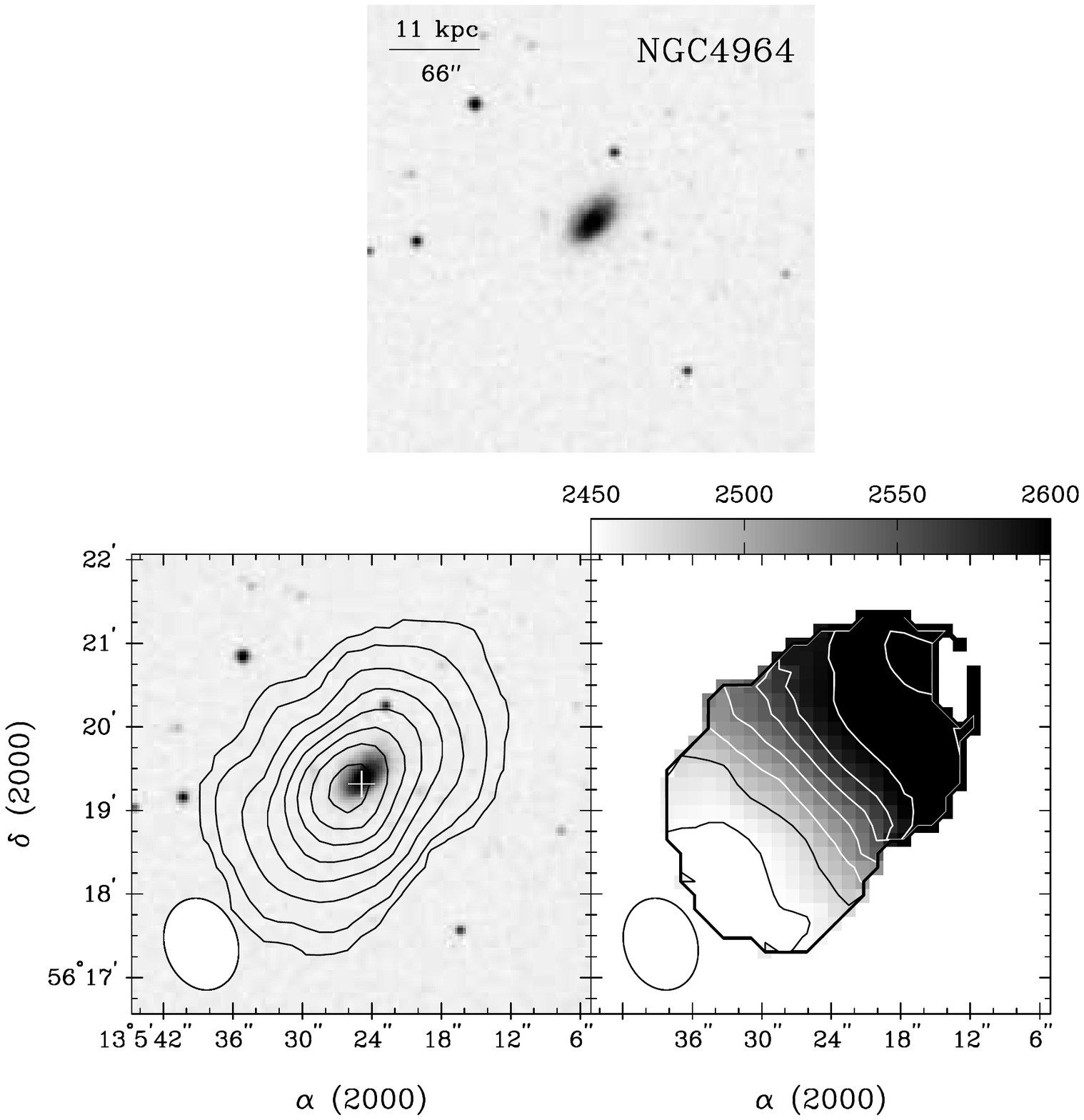}
\vspace{-2.5 cm}
\caption{Upper panel: Optical image of NGC 4964 (Control sample)
from the DSS2. Lower panels: (\textsl{Left}) Contours of zeroth
moment overlaid on the DSS2 image, and (\textsl{Right}) first
moment map. In the zeroth moment map, contours are plotted at 3,
10, 20, 30, $40 \times 29.4 {\rm \ mJy \ beam^{-1} \ km \ s^{-1}}$
($1.4 \times 10^{19} {\rm \ cm^{-2}}$). The half-power width of
the synthesized beam has a size of $67 \arcsec \times 52\arcsec$.}

\end{figure}

\clearpage
\begin{figure}
\vspace{-2 cm}
\hspace*{-0.5 cm}
\includegraphics[angle=0, scale=0.8]{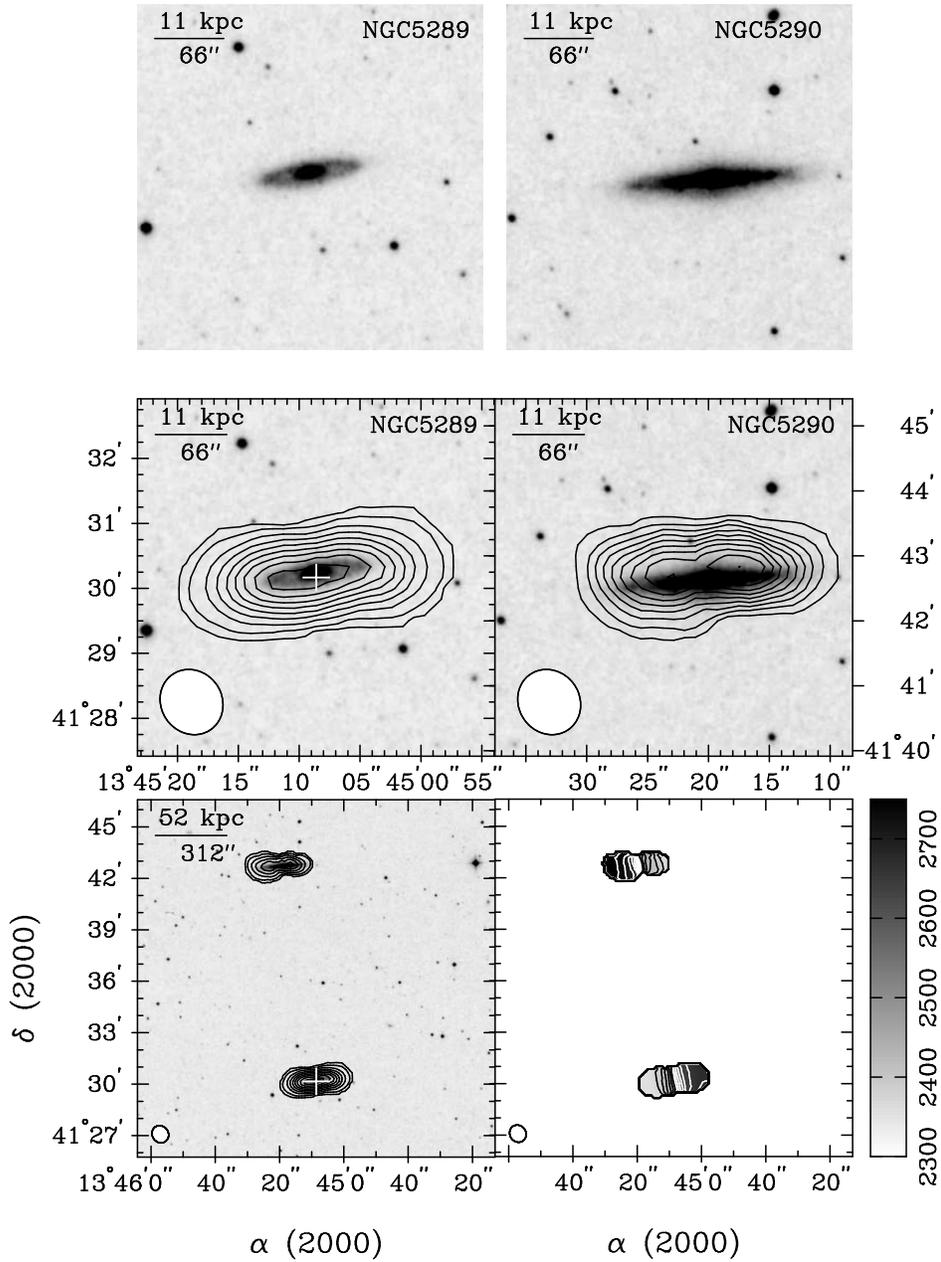}
\vspace{-1.3 cm}
\caption{Upper panel: Optical image of NGC 5289 (Control sample)
and NGC 5290 from the DSS2. Middle panels: Contours of zeroth
moment overlaid on the DSS2 image. Lower panels: (\textsl{Left})
Contours of zeroth moment overlaid on the DSS2 image and
(\textsl{Right}) first moment map with larger field. In the zeroth
moment maps, contours are plotted at 3, 10, 20, 30, $40 \times
20.8 {\rm mJy beam^{-1} km s^{-1}} ({\rm 9.6\times10^{18}
cm^{-2}})$. The half-power width of the synthesized beam has a
size of $62\arcsec \times 56 \arcsec.$}

\end{figure}

\clearpage
\begin{figure}
\vspace{-2 cm}
\hspace*{-0.5 cm}
\includegraphics[angle=0, scale=0.8]{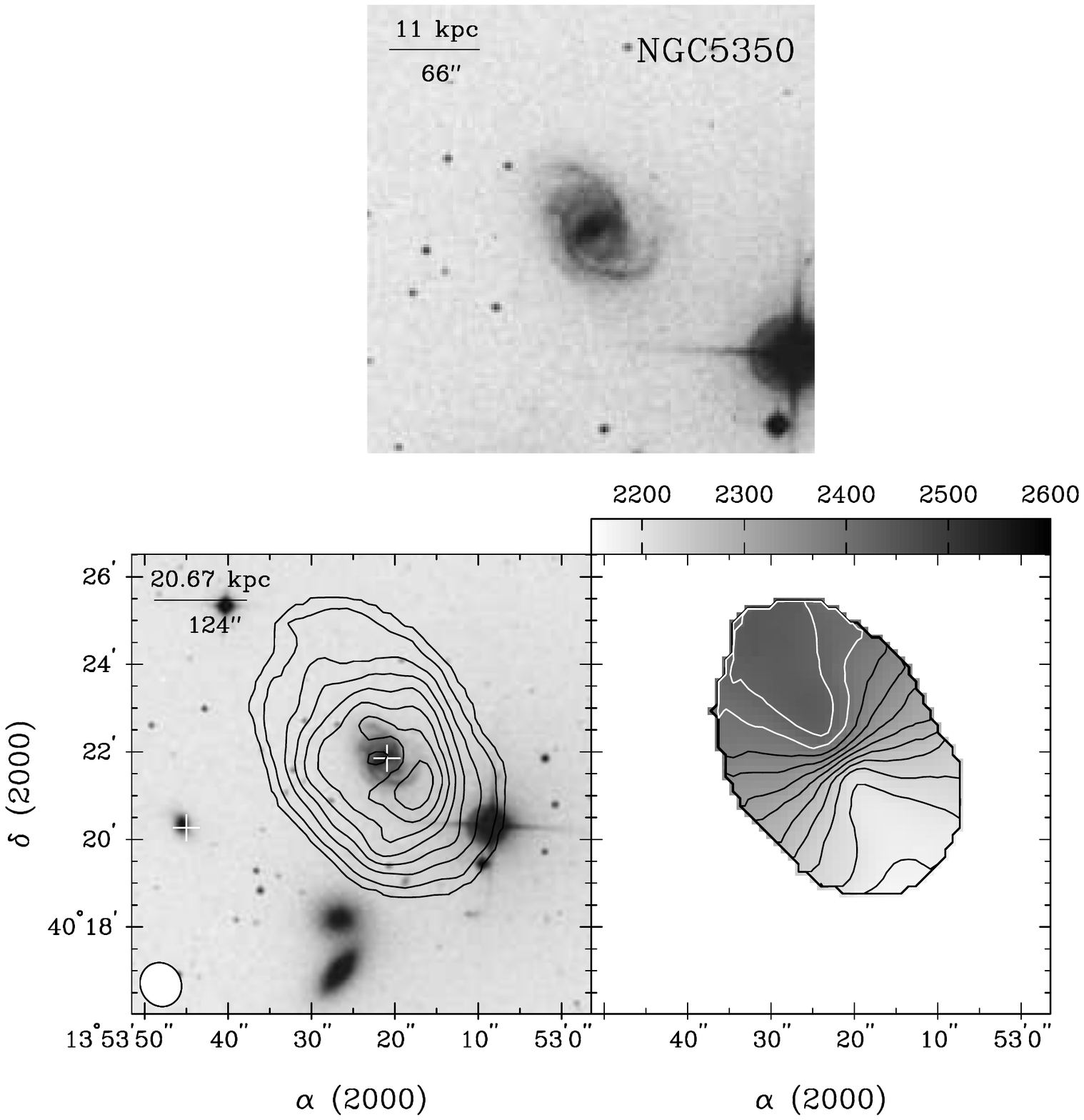}
\vspace{-2.5 cm}
\caption{Upper panel: Optical image of NGC 5350 (Control sample)
from the DSS2. Lower panels: (\textsl{Left}) Contours of zeroth
moment overlaid on the DSS2 image, and (\textsl{Right}) first
moment map. In the zeroth moment map, contours are plotted at 3,
10, 20, 30, $40 \times 31.8 {\rm \ mJy \ beam^{-1} \ km \ s^{-1}}$
($1.5 \times 10^{19} {\rm \ cm^{-2}}$). The half-power width of
the synthesized beam has a size of $61 \arcsec \times 56\arcsec$.}

\end{figure}

\clearpage
\begin{figure}
\vspace{-2 cm}
\hspace*{-0.5 cm}
\includegraphics[angle=0, scale=0.8]{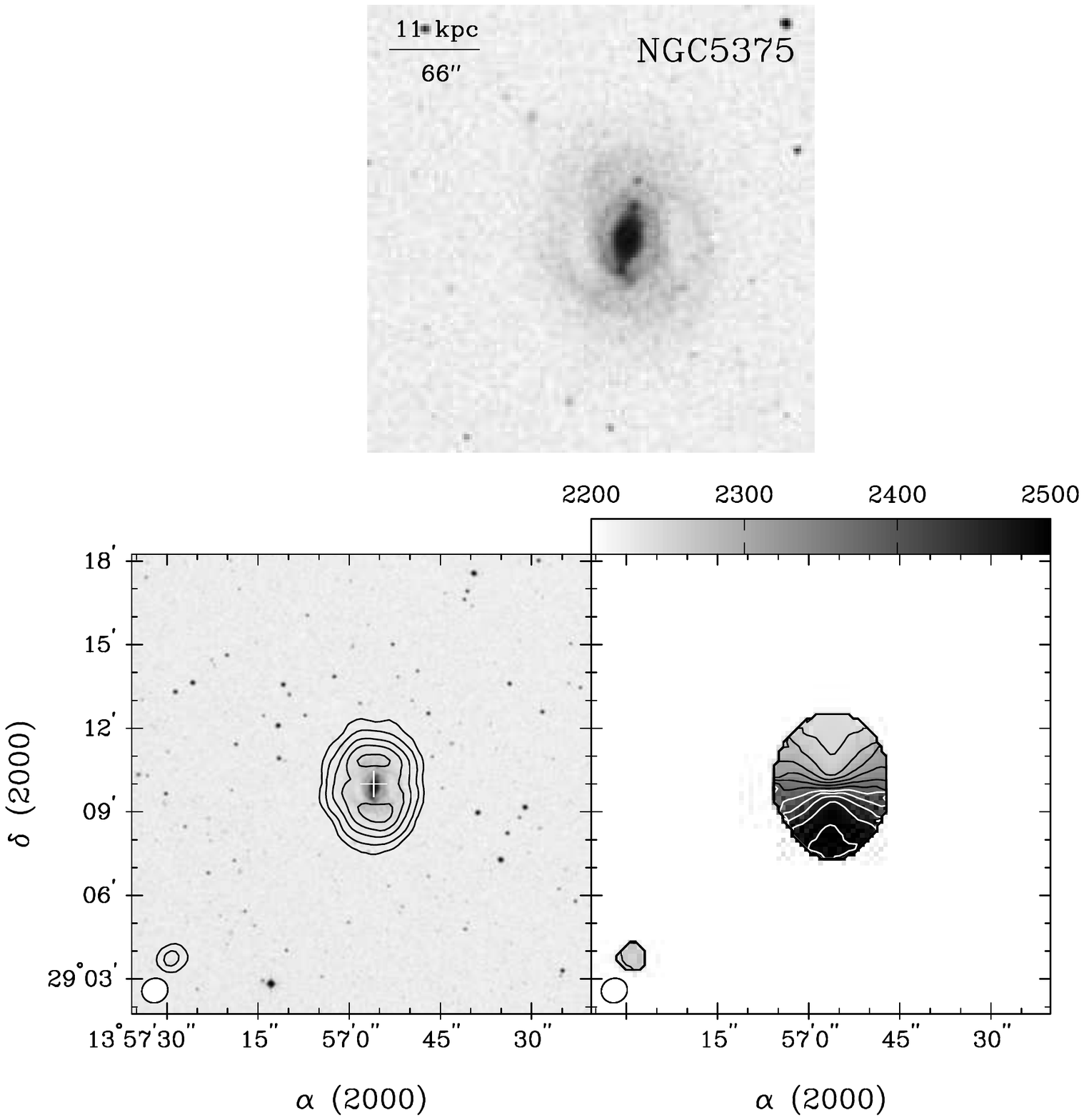}
\vspace{-2.5 cm}
\caption{Upper panel: Optical image of NGC 5375 (Control sample)
from the DSS2. Lower panels: (\textsl{Left}) Contours of zeroth
moment overlaid on the DSS2 image, and (\textsl{Right}) first
moment map. In the zeroth moment map, contours are plotted at 3,
10, 20, 30, $40 \times 32.3 {\rm \ mJy \ beam^{-1} \ km \ s^{-1}}$
($1.8 \times 10^{19} {\rm \ cm^{-2}}$). The half-power width of
the synthesized beam has a size of $57 \arcsec \times 52\arcsec$.}

\end{figure}

\clearpage
\begin{figure}
\vspace{-4.5 cm}
\hspace{-2 cm}
\includegraphics[angle=0.0, scale=0.9]{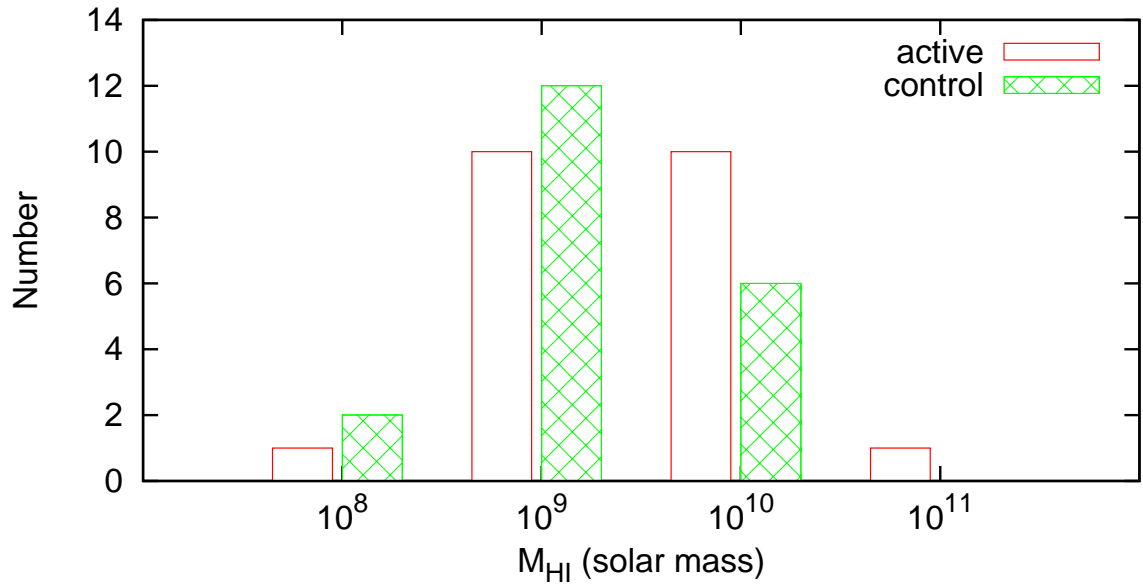}
\vspace{-1.5 cm}
\caption{Distribution in HI gas masses of ensemble active galaxy
sample of \citet{kuo08} (solid histogram) and our ensemble
control sample (hatched histogram).  The combined HI gas masses of
NGC 4567/NGC 4568 in the control sample is plotted.}
\end{figure}

\clearpage
\begin{figure}
\vspace{-14 cm}
\hspace{0.0 cm}
\includegraphics[angle=0, scale=1.0]{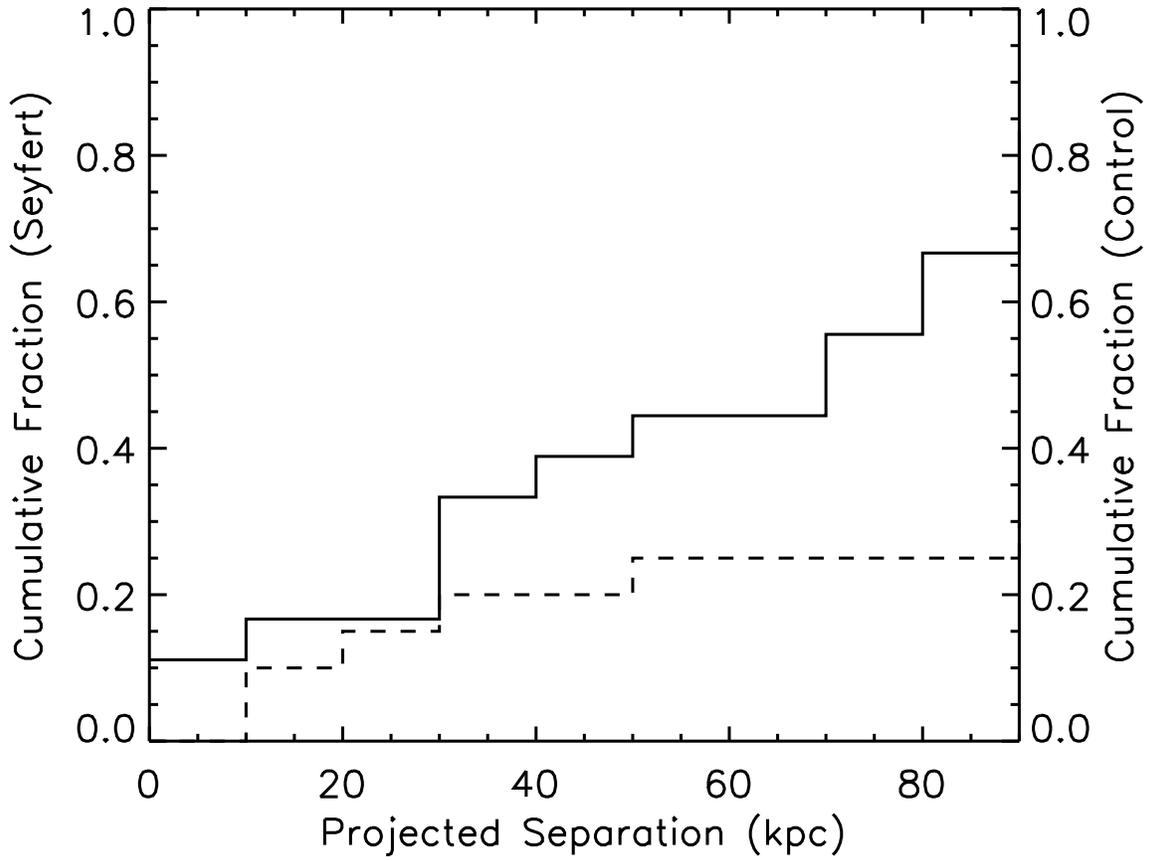}
\vspace{0.2 cm}
\caption{Cumulative fraction of Seyfert galaxies (solid line) and
control sample (dashed line) with interacting neighboring galaxies
plotted as a function of their projected separations.}
\end{figure}

\clearpage


\newpage
\begin{thebibliography}{}

\bibitem[Condon et al.(1998)]{con98} Condon, J. J., Cotton, W. D., Greisen, E. W., Yin, Q. F., Perley, R. A., Taylor, G. B., \& Broderick, J. J., \aj, 115, 1693


\bibitem[Dahari(1984)]{dah84}Dahari, O. 1984, \aj, 89, 966

\bibitem[de Robertis, Hayhoe \& Yee(1998)]{der98a} de Robertis, M. M., Hayhoe, K., \& Yee, H. K. C. 1998a, \apj, 115, 163

\bibitem[de Robertis, Yee, \& Hayhoe(1998)]{der98b} de Robertis, M. M., Yee, H. K. C., \& Hayhoe, K. 1998b, \apj, 496, 93

\bibitem[Dultzin-Hacyan et al.(1999)]{dul99}Dultzin-Hacyan, D., Krongold, Y., Fuentes-Guridi, I., \& Marziani, P. 1999, \apj, 513, 111

\bibitem[Fuentes-Williams \& Stocke(1988)]{fue88}Fuentes-Williams, T. \& Stocke, J. T. 1988, \aj, 96, 1235

\bibitem[Giovanelli \& Haynes(1981)]{Giovanelli1981} Giovanelli, R. \&
Haynes, M. 1981, \aj, 86, 340

\bibitem[Huchra et al.(1983)]{huc83}Huchra, J., Davis, M., Latham, D., \& Tonry, J. 1983, \apjs, 52, 89

\bibitem[Huchra et al.(1995)]{huc95}Huchra, J. P., Geller, M. J., \& Corwin, H. G. Jr. 1995, \apjs, 99, 391

\bibitem[Huchtmeier (1989)]{Huchtmeier1989} Huchtmeier, W. K. \& Richter,
O.-G. 1989, A General Catalog of H I Observations of Galaxies
(Berlin: Springer)

\bibitem[Iono et al.(2005)]{Iono2005} Iono, D., Yun, M. S., \& Ho,
P.T.P. 2005, \apjs, 158, 1

\bibitem[Koulouridis et al.(2006)]{kou06}Koulouridis, E., Plionis, M., Chavushyan, V., Dultzin-Hacyan, D., Krongold, Y., \& Goudis, C. 2006, \apj, 639, 37

\bibitem[Kuo et al.(2008)]{kuo08} Kuo, C.-Y., Lim, J., Tang, Y.-W., \& Ho, P. T. P. 2008, submitted

\bibitem[Lipovetsky et al.(1988)]{lip88}Lipovetsky, V. A., Neizvestny, S. I., \& Neizvestnaya, O. M. 1988, Soob. Spets. Astrofiz. Obs., 55, 5

\bibitem[MacKenty(1989)]{mac89}MacKenty, J. W. 1989, \apj, 343, 125

\bibitem[Martini(2004)]{mart04}Martini, P. 2004, in IAU Symp. 222, The Interplay among Black Holes, Stars and ISM in Galactic Nuclei, ed. Th. S. Bergmann, L. C. Ho, \& H. R. Schmitt (San Francisco: ASP), 557

\bibitem[Paturel et al.(2003)]{pat03} Paturel, G., Petit, C., Prugniel, Ph., Theureau, G., Rousseau, J., Brouty, M., Dubois, P., \& Cambr\'esy, L. 2003, \aap, 412, 45

\bibitem[Rafanelli et al.(1995)]{raf95}Rafanelli, P., Violato, M., \& Baruffolo, A. 1995, \aj, 109, 1546

\bibitem[Schmitt(2001)]{sch01}Schmitt, H. R. 2001, \aj, 122, 2243

\bibitem[Stauffer(1982)]{sta82}Stauffer, J. R. 1982, \apj, 262, 66

\bibitem[V\'eron-Cetty \& V\'eron(1998)]{ver98} V\'eron-Cetty, M.-P. \& V\'eron, P. 1998, A Catalogue of Quasars and Active Nuclei (8th ed.; Garching: ESO)

\bibitem[V\'eron-Cetty \& V\'eron(2000)]{ver00}V\'eron-Cetty, M.-P. \& V\'eron, P. 1998, A Catalogue of Quasars and Active Nuclei (9th ed.; Garching: ESO)

\bibitem[Veron-Cetty \& Veron(2006)]{ver06} V\'eron-Cetty, M.-P. \& V\'eron, P. 2006, \aap, 455, 773

\bibitem[Yun, Ho, \& Lo(1994)]{yun94}Yun, M. S., Ho, P. T. P., \& Lo, K. Y. 1994, \nat, 372, 530

\end{thebibliography}
\end{document}